\documentclass[final,3p,times]{elsarticleNew}

\usepackage{amssymb}

\usepackage{lineno}

\usepackage{amsmath}
\usepackage{url}
\usepackage[hidelinks]{hyperref}
\usepackage{comment}
\biboptions{sort&compress}

\journal{Computers \& Fluids}

\begin{document}

\begin{frontmatter}



\title{Wall-modeled large-eddy simulation of turbulent smooth body separation using the OpenFOAM flow solver}


\author[inst1]{Christoffer Hansen}

\author[inst2]{Xiang I. A. Yang}

\author[inst1]{Mahdi Abkar\corref{cor1}}
\cortext[cor1]{Corresponding author}
\ead {abkar@mpe.au.dk}

\affiliation[inst1]{organization={Department of Mechanical and Production Engineering, Aarhus University},
            city={Aarhus N},
            postcode={8200},
            country={Denmark}}

\affiliation[inst2]{organization={Department of Mechanical Engineering, Pennsylvania State University},
            city={State College},
            postcode={16802}, 
            state={PA},
            country={USA}}

\begin{abstract}
This work investigates the current wall-modeled large-eddy simulation (WMLES) capabilities of the open-source computational fluid dynamics solver OpenFOAM, which is used widely in academia and industry.
This is achieved by a simulation campaign that covers both attached and smooth body separation cases.
The campaign includes simulations using four different wall models and aims to investigate the sensitivity of the results to changes in numerics, mesh resolution, and subgrid-scale modeling.
The results demonstrate that two main factors largely determine OpenFOAM-based WMLES performance.
These are the discretization of the convective term and wall modeling.
For the former, the best performance in the attached case is achieved with low-dissipation numerics,
however, for the smooth body separation case, more dissipative numerics give the best performance. 
For the latter, we find that both equilibrium and non-equilibrium wall models perform well in the attached case but that the non-equilibrium models significantly improve the prediction of smooth body separation.
Still, the non-equilibrium wall model results do not show a uniform improvement over equilibrium models.
This is explained by an inconsistent accounting of non-equilibrium physics in these models, i.e., including the pressure gradient term without also including the convective term.
This highlights the potential for future performance improvements by using non-equilibrium wall models that consistently account for both the convective and pressure gradient terms.
\end{abstract}

\begin{keyword}
wall modeling \sep large-eddy simulation \sep OpenFOAM \sep smooth body separation
\end{keyword}

\end{frontmatter}




\section{Introduction}
\label{sec:introduction}

With the continued growth of computational power, scale-resolving simulations of complex turbulent flows are becoming increasingly common.
An example of this trend is the recent appearance of scale-resolving simulations of full aircraft, e.g., \citep{goc2021large}, which would have been infeasible only a decade ago.
Still, given the large computational cost associated with direct numerical simulations (DNS) of high Reynolds number flows \citep{moin1998direct}, some degree of modeling is still needed to reduce the computational cost.
One of the most promising scale-resolving methodologies for achieving this is large-eddy simulations (LES).
In LES, the computational cost is reduced by only solving for the large-scale motions directly, while the effects of the small-scale motions are included using a model \citep{rogallo1984numerical, lesieur1996new}.
As small-scale motions account for most of the computational expense in DNS, the computational savings of LES compared to DNS can be significant.
However, for wall-bounded turbulent flows, even the largest scales within the near-wall region are very small at high Reynolds numbers.
Therefore, if LES is to provide significant computational savings over DNS for wall-bounded flows, additional modeling of the near-wall flow is needed to allow a looser grid in this region \citep{chapman1979computational, choi2012grid, yang2021grid, agrawal2023grid}.
This approach is called wall-modeled large-eddy simulations (WMLES) and was pioneered around 50 years ago \citep{deardorff1970numerical, schumann1975subgrid}.
Since then, it has seen continuous interest and development as documented in several reviews \citep{piomelli2002wall, larsson2016large, bose2018wall}.

As WMLES continues to move from an academic to an applied setting, the need for careful validation becomes increasingly important.
This is especially true for practitioners, who need solid knowledge about the expected accuracy of WMLES for their respective cases of interest.
Similarly, validation is essential in reducing the many possible combinations of different meshes, numerics, and modeling into a more manageable set of best practices.
Validation in complex geometries also helps identify the remaining limitations of current turbulence models, thereby focusing future developments in WMLES modeling where they matter most.
This crucial need for validation has resulted in several important validation initiatives.
This includes the High Lift Prediction Workshop \citep{ashton2024summary} and the WMLES component of the High-Fidelity Computational Fluid Dynamics (CFD) Verification Workshop \citep{garmann2024summary}.
Another example is the recent Virginia Tech and NASA CFD Turbulence Model Validation Challenge focused on the BeVERLI (Benchmark Validation Experiments for RANS and LES Investigations) Hill geometry \citep{roy2023blind, lowe2024experimental, roy2024summary}.
Several recent validation efforts also exist in the general literature.
One such example is \citep{li2025analysis}, which investigated the WMLES capabilities of ANSYS FLUENT in turbulent channels.
Further, the performance of WMLES using several different machine-learning-based wall models was recently survived in \citep{vadrot2023survey}.
Finally, we note that a better understanding of grid convergence in WMLES is also needed to properly validate it to industry standards.
Recent work on grid convergence in WMLES can be found in \citep{hu2024grid, yang2024grid} and references therein.

In the present work, we focus on WMLES using the open-source, finite-volume code OpenFOAM \citep{weller1998tensorial}, specifically the OpenFOAM variant developed and maintained by OpenCFD Limited.
While OpenFOAM does support WMLES, the out-of-the-box selection of wall models is quite limited, and matching with the outer LES is only possible in the first off-wall cell center.
Therefore, to enhance the WMLES capabilities of OpenFOAM, Mukha et al. \citep{mukha2019library} recently introduced a dedicated WMLES library. The library provides a broader selection of wall models and increased flexibility in picking the matching location, along with other improvements.
In the original article \citep{mukha2019library}, the library was tested for WMLES of turbulent channels and flow over a backward-facing step.
The authors tested the performance of the implemented wall models using first— and second-point matching, different mesh resolutions, and numerical discretizations of the convective term.
They considered a single subgrid-scale (SGS) model.
An interesting finding was that the non-equilibrium wall models, which include pressure gradient effects, performed worse than the equilibrium models on the backward-facing step case.
The library has also been used for uncertainty quantification of WMLES in turbulent channels \citep{rezaeiravesh2019systematic}.
An equilibrium wall model was used, and three different SGS models were considered.
Further, the library has been tested on unstructured meshes in channels and zero-pressure-gradient boundary layers \citep{mukha2021predictive}.
An equilibrium wall model was used, and a single SGS model was considered.
While these works have laid a good foundation regarding the validation of OpenFOAM-based WMLES using the library from \citep{mukha2019library}, one shortcoming is the primary focus on simple attached flows without the presence of non-equilibrium effects and smooth body flow separation.
Therefore, to further validate the capabilities and robustness of OpenFOAM-based WMLES using the library from \citep{mukha2019library}, we consider a periodic hill case, which features smooth body separation and strong non-equilibrium effects.
The case was first introduced in \citep{mellen2000large} and has since become a popular simulation benchmark in the turbulence community \citep{frohlich2005highly,breuer2009flow,duprat2011wall,gloerfelt2019large,bin2023large}.
To properly benchmark the WMLES performance, we consider a broad simulation campaign that covers different wall models, SGS models, numerics, and mesh resolutions.
To give further insight into the performance of the different wall models, we also carry out an \textit{a-priori} analysis using reference data from a wall-resolved LES (WRLES) of the periodic hill case.
Finally, we note that only matching at the first off-wall cell center is considered in this work.
This choice was made because the primary focus is placed on the periodic hill case, and preliminary investigations showed that matching at the second or third points degraded performance.

The rest of this paper is organized as follows: In Section \ref{sec:LES}, we review the LES formalism, including both SGS and wall modeling.
Next, we discuss the simulation details (solvers, discretizations, etc.) and provide an overview of the different simulation cases in Section \ref{sec:sim_details}.
The results of the simulations are then presented in Section \ref{sec:results}.
Finally, in Section \ref{sec:conclusions}, we provide some concluding remarks and suggestions for future developments.

\section{Large-eddy simulations}
\label{sec:LES}

This section provides background information on the LES formalism.
We first briefly cover the motivation and development of the method in Section \ref{subsec:LES_motiv_and_devel}.
In Section \ref{subsec:SGS_models}, we review SGS modeling focusing on the Wall-Adapting Local Eddy-viscosity (WALE) and Dynamic Smagorinsky (DS) models.
We then review wall modeling in Section \ref{subsec:wall_models} and provide details on the specific wall models considered.

\subsection{Motivation and development}
\label{subsec:LES_motiv_and_devel}

Motivated by reducing the high computational cost of DNS, LES aims to solve for the large-scale motions and accounts for the effect of smaller scales using an appropriate model. 
To obtain governing equations for the large-scale motions, Leonard \citep{leonard1975energy} suggested applying a low-pass filter to the Navier-Stokes equations.
This filtering introduces new terms in the LES governing equations, which depend on the unknown small-scale motions, causing the LES closure problem.
Following common practice, we will assume that the low-pass filter is homogeneous.
This assumption ensures that the filtering and differentiation operations commute, simplifying the LES governing equations.
For additional details and a discussion of the non-homogeneous case, we refer to \citep{ghosal1995basic,sagaut2005large,yalla2021effects}.
Another important question is the choice of a filtering length scale.
Traditionally, the filtering length scale is chosen to be within the inertial subrange.
While this is reasonable for simple cases, such as homogeneous isotropic turbulence or turbulent free-shear flows, it becomes problematic for wall-bounded turbulent flows.
The reason is that the length scales associated with the inertial range in a near-wall part of a flow are much smaller than for parts of the flow that are far from the walls.
This necessitates additional modeling of the small-scale motions in the near-wall parts of the flow if LES is to remain significantly cheaper than DNS.
For further details on the historical development and philosophy of LES, we refer to \citep{rogallo1984numerical,lesieur1996new,meneveau2000scale} and references therein.

In this work, we consider the filtered incompressible Navier-Stokes equations written in index notation (note summation over repeated indices)
\begin{align}
    \partial_j \tilde{u}_j &= 0 , \\
    \partial_t \tilde{u}_i + \partial_j ( \tilde{u}_i \tilde{u}_j ) &= - \partial_i \tilde{p} + \nu \partial_j \partial_j \tilde{u}_i - \partial_j \tau_{ij} ,
\end{align}
where $\partial_t = \partial / \partial t$ and $\partial_i = \partial / \partial x_i$.
Here $x_i$ and $u_i$ for $i = 1,2,3$ (or interchangeably $x,y,z$ and $u,v,w$) are the streamwise, wall-normal, and spanwise coordinates and velocity components, respectively, and $\tilde{\cdot}$ indicates filtered variables.
Further, $p$ is the modified (density weighted) pressure, $\nu$ is the kinematic viscosity, and $\tau_{ij}$ is the SGS tensor, given by
\begin{equation}
    \tau_{ij} = \widetilde{u_i u_j} - \tilde{u}_i \tilde{u}_j .
\end{equation}
\subsection{Subgrid-scale modeling}
\label{subsec:SGS_models}
The goal of SGS modeling is to account for the effect of small-scale motions that are not resolved on the computational grid.
The most popular approach to SGS modeling is using eddy-viscosity models based on the Boussinesq hypothesis \citep{meneveau2000scale}.
In this case, the deviatoric part of the SGS tensor, denoted as $\tau_{ij}^\Delta$, is modeled as
\begin{align}
\label{eq:eddy_viscosity_model}
    \tau_{ij}^\Delta = \tau_{ij} - \frac{1}{3} \tau_{kk} \delta_{ij} = -2 \nu_t \tilde{S}_{ij} ,
\end{align}
while the isotropic part is included in the pressure gradient term.
Here $\nu_t$ is the turbulent eddy-viscosity, and $\tilde{S}_{ij}$ is the filtered rate-of-strain tensor defined as
\begin{align}
    \tilde{S}_{ij} = \frac{1}{2} \left( \partial_j \tilde{u}_i + \partial_i \tilde{u}_j \right) .
\end{align}
We will consider two different SGS models in this work.
The first is the Wall-Adapting Local Eddy-viscosity (WALE) model \citep{nicoud1999subgrid}.
The second is the Dynamic Smagorinsky (DS) model \citep{germano1991dynamic} with the modification introduced by Lilly \citep{lilly1992proposed}.
We chose these as they are two of the most extensively used SGS models.

In the WALE model, the turbulent eddy-viscosity is expressed in terms of the square of the velocity gradient tensor.
Denoting the filtered velocity gradient tensor as $\tilde{g}_{ij} = \partial_j \tilde{u}_i$, the traceless symmetric part of the square of the velocity gradient tensor is
\begin{align}
    \tilde{\xi}_{ij} = \frac{1}{2} \left( \tilde{g}_{ij}^2 + \tilde{g}_{ji}^2 \right) - \frac{1}{3} \tilde{g}_{kk}^2 ,
\end{align}
where $\tilde{g}_{ij}^2 = \tilde{g}_{ik}\tilde{g}_{kj}$.
We can then write the turbulent eddy-viscosity for the WALE model as
\begin{align}
    \nu_t^{\text{WALE}} = ( C_w \Delta )^2 \frac{( \tilde{\xi}_{ij} \tilde{\xi}_{ij} )^{3/2}}{( \tilde{S}_{ij} \tilde{S}_{ij} )^{5/2} + \left( \tilde{\xi}_{ij} \tilde{\xi}_{ij} \right)^{5/4}} .
\end{align}
In the WALE model, $C_w$ is considered a true constant, and thus, it is not tuned using a dynamic procedure.

The turbulent eddy-viscosity for the DS model is expressed in terms of the filtered rate-of-strain tensor and has the same form as in the original Smagorinsky model
\begin{align}
    \nu_t^{\text{DS}} = (C_s \Delta )^2 (2 \tilde{S}_{ij} \tilde{S}_{ij})^{1/2} .
\end{align}
However, contrary to the original Smagorinsky and WALE models, in the DS model, the coefficient $C_s$ is tuned locally in the flow using a dynamic procedure based on the Germano identity \citep{germano1992turbulence}.
For further details on the dynamic procedure, we refer to \citep{germano1991dynamic, lilly1992proposed}.

While the WALE model is part of the standard OpenFOAM catalog of SGS models, this is not true for the DS model.
For this reason, we use the WALE model as the primary SGS model in this work, while the DS model is included to investigate the sensitivity of the results to SGS modeling.
Regarding the DS model implementation, we use the library by Alberto Passalacqua \citep{DS_lib_OpenFOAM}.

\subsection{Wall models}
\label{subsec:wall_models}

Similar to SGS models, wall models account for the effects of unresolved small-scale motions in the near-wall flow.
While wall modeling can be approached in several ways, in this work, we focus on wall stress modeling as defined in \citep{larsson2016large}.
In wall stress modeling, the traditional no-slip boundary condition is replaced by an approximate Neumann boundary condition, which can be directly related to the wall shear stress.
Therefore, in this context, wall modeling reduces to the task of reconstruction of the flow field in the wall layer using information from the near-wall cell centers \citep{lv2021wall,hansen2023pod} and specifying an appropriate wall shear stress accordingly.

The most widely used type of wall model is the equilibrium wall model (EWM) \citep{schumann1975subgrid,bou2005scale,kawai2012wall,de2021unified}.
The algebraic variant of EWMs computes the wall shear stress according to some mean flow scaling in the wall-adjacent computational cell(s), usually the law of the wall (LoW).
The mean flow scaling is matched with the LES velocity at an off-wall location locally and instantaneously, resulting in algebraic equations for the wall shear stress.
A popular variant is based on solving the thin boundary-layer equations (TBLEs).
In these EWMs, only the Reynolds stress and viscous stress terms are retained, and the Reynolds stress term is closed using a zero-equation eddy-viscosity model \citep{kawai2012wall,yang2018semi,chen2022wall}.
The resulting ordinary differential equation (ODE) is then solved on a fine near-wall grid using the no-slip condition at the wall.
The off-wall boundary condition is provided by matching with the LES in the wall-adjacent cell(s).
The wall shear stress can then be calculated from the ODE solution.
Although EWMs are favorable regarding their simplicity and stability, it has long been known that EWMs can struggle in non-equilibrium flows \citep{piomelli2002wall,larsson2016large,bose2018wall}. 
This has prompted many attempts at creating more general non-equilibrium wall models.
One approach is based on solving the full TBLEs
\begin{equation}
    \partial_2 \left[ \left( \nu + \nu_t \right) \partial_2 \left\langle u_i \right\rangle \right] = F_i ,
\label{eq:TBLE}
\end{equation}
where
\begin{equation}
    F_i = \partial_t \langle u_i \rangle + \partial_j ( \langle u_i \rangle \langle u_j \rangle ) + \partial_i \langle p \rangle ,
\end{equation}
with all of the non-equilibrium terms included \citep{park2014improved}. 
However, as this involves solving an additional system of partial differential equations, it comes with a significant computational overhead \citep{park2016numerical}.
The solution procedure is also complicated by the need for a separate connected near-wall mesh, which is especially challenging in complex geometries.
One way of overcoming this is by taking an integral approach combined with an assumed analytical form for the LES-grid filtered velocity as done in the integral wall model \citep{yang2015integral} and the Lagrangian relaxation towards equilibrium wall model \citep{fowler2022lagrangian}.
This results in wall models that account for non-equilibrium effects while only requiring the solution of algebraic equations.
Another approach to move beyond the limitations of EWMs is to mathematically derive a slip boundary condition from the filtered Navier-Stokes equations.
This slip boundary condition is used instead of the more traditional Neumann boundary condition.
The resulting wall model is called the dynamic slip model \citep{bose2014dynamic,bae2019dynamic}.
With the rapidly growing interest in machine learning for fluid dynamics and turbulence modeling \citep{kutz2017deep,wu2018physics,duraisamy2019turbulence,brunton2020machine,beck2021perspective}, several machine-learning-based wall models have also been proposed in recent years.
Most of these wall models are based on supervised learning, where a neural network is trained using DNS data to predict the wall shear stress from information at the near-wall cell center(s)
\citep{yang2019predictive,zhou2021wall}.
A promising variation of this approach is the building-block-flow wall model introduced in \citep{lozano2023machine}.
This model's main assumption is that the physics of complex flows can be seen as a combination of flow phenomena from a set of canonical turbulent flows. 
The model combines a neural network for classifying the building block flows and a prediction network that estimates the wall shear stress using the classifier's output and local flow information. 
Recently, modal-based non-equilibrium wall models have also been introduced \citep{hansen2023pod, hansen2024extension}.
These models are similar to algebraic EWMs, but the LoW is augmented by additional degrees of freedom derived from modal analysis, allowing for a more accurate parametrization of the near-wall velocity.
Multiple matching locations are used to tune the additional degrees of freedom.
Finally, wall models have also been developed using reinforcement learning \citep{bae2022scientific,vadrot2023log}.

In this work, we consider a class of wall models called ODE models \citep{balaras1996two, wang2002dynamic, duprat2011wall, kamogawa2023ordinary}.
These models consider simplifications of the TBLEs, Eq. \eqref{eq:TBLE}, resulting in approximate ODEs.
Here we focus on the cases when $F_i = 0$ (EWM) and $F_i = \partial_i \langle p \rangle$.
Note that the pressure gradient is assumed constant and determined by matching with the outer LES solution.
For these cases, a closed-form solution for the wall shear stress can be derived \citep{wang2002dynamic}.
To see this, we start by integrating Eq. \eqref{eq:TBLE} from the wall up to a given height $y$, which gives
\begin{align}
    (\nu + \nu_t) \partial_y \langle u_i \rangle - \frac{\langle \tau_{w,i} \rangle}{\rho} = F_i y,
\end{align}
where we have used that $F_i$ is assumed constant.
Using this and rewriting, we get
\begin{align}
    \frac{(\langle \tau_{w,i} \rangle / \rho)}{\nu + \nu_t} = \partial_y \langle u_i \rangle - F_i \frac{y}{\nu + \nu_t} .
\end{align}
Next, we integrate from $y=0$ to $y=h$, with $h$ being the wall-normal distance to the matching location.
Taking $( \langle \tau_{w,i} \rangle / \rho)$ outside of the integral on the left-hand side, applying the no-slip condition in the lower limit for the velocity term, and rearranging slightly, we arrive at
\begin{equation}
\label{eq:tauw_formula}
    \frac{\langle \tau_{w,i} \rangle}{\rho} = \frac{1}{\int_0^h \frac{dy}{\nu + \nu_t}} \left[ \langle u_i \rangle|_{y=h} - F_i \int_0^h \frac{y \, dy}{\nu + \nu_t}  \right] .
\end{equation}
As this equation is typically implicit, due to $\nu_t$ depending on $\langle \tau_{w,i} \rangle$ through the friction velocity $u_\tau = \sqrt{\langle \tau_w \rangle / \rho}$ where $\langle \tau_w \rangle$ is the wall shear stress magnitude, it is solved using and iterative method.

For the eddy-viscosity, we consider two different models.
Both use a damped mixing-length closure, see, e.g., \citep{cabot2000approximate}, but they differ in the specific damping function being used.
The first models the eddy-viscosity using the van Driest damping function \citep{van1956turbulent}.
The expression for the eddy-viscosity is
\begin{equation}
    \nu_t = \nu \kappa y^+ \left( 1 - \exp(-y^+/A) \right)^2 ,
\label{eq:van_Driest_eddy_viscosity}
\end{equation}
where $y^+ = u_\tau y / \nu$ is the wall-unit-scaled distance from the wall, and we use $\kappa = 0.4$ and $A = 17.8$ following \citep{mukha2019library}.
The second eddy-viscosity model was proposed by Duprat et al. \citep{duprat2011wall}, and tries to account for pressure gradient effects.
The model uses the combined velocity scale $u_{\tau,p} = (u_\tau^2 + u_p^2)^{1/2}$, where $u_p = [\nu / \rho (dp/dx)]^{1/3}$ is the pressure based velocity scale introduced in \citep{manhart2008near}.
We note $u_p$ has recently gained increased attention, both as a feature in data-driven wall modeling \citep{zhou2021wall,zhou2023wall,zhou2024wall} and for constructing sensors which allow switching between different wall models in equilibrium and non-equilibrium regions \citep{agrawal2022wall,agrawal2024non}.
The non-dimensional parameter $\alpha = u_\tau^2 / u_{\tau,p}^2$ is also introduced to measure the balance between shear stress and the streamwise pressure gradient.
Specifically, $\alpha = 0$ corresponds to a zero shear stress flow, i.e., a separation point, and $\alpha = 1$ corresponds to a zero pressure gradient flow.
The model expression is
\begin{equation}
    \nu_t = \nu k y^* \left( \alpha + y^*(1 - \alpha)^{3/2} \right)^\beta \left[ 1 - \exp\left( -\frac{y^*}{1 + A \alpha^3} \right) \right]^2 ,
\label{eq:Duprat_eddy_viscosity}
\end{equation}
where $y^* = u_{\tau,p} y / \nu$ is the wall-distance scaled using the combined velocity scale.
Further, we use $\kappa = 0.4$, $A = 17$, and $\beta = 0.78$ following \citep{duprat2011wall}.

In OpenFOAM, the wall shear stress from the wall model is enforced by specifying a non-zero eddy-viscosity at the wall \citep{mukha2019library}.
To see how this works, consider the calculation of the wall shear stress in the finite-volume method
\begin{align}
    \tilde{\tau}_{w,i} \approx (\nu + \nu_t) \frac{\tilde{u}_{i,h}}{h} , \qquad i=1,3,
\end{align}
where $\tilde{u}_{i,h} = \tilde{u}_i(y=h)$ is the LES velocity at the matching location.
Therefore, to enforce a given wall shear stress magnitude $\tilde{\tau}_w$, the eddy-viscosity should be specified as
\begin{align}
    \nu_t = \frac{\tilde{\tau}_w}{\left[ (\tilde{u}_{1,h} / h)^2 + (\tilde{u}_{3,h} / h)^2  
 \right]^{1/2}} - \nu .
\end{align}
We note that this approach limits the wall shear stress to be aligned with the wall-parallel velocity at the first off-wall cell center, which is not necessarily true for fully three-dimensional boundary layers.

Before moving on, we want to comment on the fact that the convective term is not included in the ODE wall models with $F_i = \partial_i \langle p \rangle$.
This topic has been discussed in several works \citep{hickel2013parametrized, larsson2016large, kamogawa2023ordinary}, and it is generally argued that including only one of the convective and pressure terms is inconsistent.
In \citep{hickel2013parametrized}, they investigated this empirically using WRLES data from two different non-equilibrium flows involving separation.
The first is a turbulent boundary layer with an imposed adverse pressure gradient \citep{hickel2008implicit}, and the second is a shock/turbulent-boundary-layer interaction \citep{touber2009large}.
They observed that the convective and pressure gradient terms approximately balance each other in the outer part of the boundary layer for both cases.
However, this balance was not observed inside the viscous layer ($y^+ \lesssim
 30 - 50$).
The approximate balance between the convective and pressure gradient terms is discussed further in \citep{larsson2016large} and used as a possible explanation of why EWMs sometimes perform well even when strong non-equilibrium effects are present.
Similarly, \citep{kamogawa2023ordinary} considered the balance of convective and pressure gradient terms in a pressure-induced separated and reattached turbulent boundary layer.
They also found that the terms almost balance at a wall-normal distance of roughly $10\%$ of the boundary layer thickness.
While these previous observations seem to establish a clear trend, it should be noted that they all come from statistically two-dimensional flows over flat plates.
Therefore, this balance should be investigated further in flows with curved surfaces and for fully three-dimensional cases.
Another point that is often overlooked in this discussion is the difference between external flows and internal pressure-driven flows.
For the latter, a balance between the convective and pressure gradient terms is obviously not expected in general.
Further, for cases such as equilibrium channel and pipe flows, including the pressure gradient without the convective term is clearly justified.
However, including only the pressure gradient might not be justified in non-equilibrium internal flows.
In this work, we investigate this question for the periodic hill configuration as discussed in Section \ref{subsec:perhill_results}.

\section{Simulation details}
\label{sec:sim_details}

We have carried out a simulation campaign consisting of two different turbulent flows.
The first is a turbulent channel flow, and the second is a turbulent flow over periodic hills.
The campaign aims to provide insight into the interplay between numerics, SGS models, and wall modeling.
Therefore, the campaign includes a large number of cases, which vary in each of these components.
Below, we give an overview of the setup for the two cases, including numerics, meshing, and relevant SGS and wall modeling details.
We note that all simulations presented use the open-source, finite-volume code OpenFOAM.
Originally developed as an in-house research code \citep{weller1998tensorial}, OpenFOAM has grown into a general-purpose CFD software suite that is widely used in both academia and industry.
The OpenFOAM variant used here is the one developed and maintained by OpenCFD Limited and distributed through the website: \url{www.openfoam.com}.
The specific OpenFOAM version used is v2312.

We first give an overview of the numerics used in the simulations, which are summarized in Table \ref{tab:overview_numerics}.
These numerics are inspired by previous efforts in OpenFOAM-based WRLES and WMLES \citep{matai2019large, mukha2019library}.
For all cases, the unsteady, incompressible solver pisoFoam is used, which utilizes the PISO algorithm (Pressure-Implicit with Splitting of Operators) developed by Issa and coworkers \citep{issa1986solution, issa1991solution}.
Second-order-accurate schemes are used for both spatial and temporal discretizations.
Specifically, gradients and Laplacians are discretized using the Gauss linear (centered) scheme, while surface-normal gradients are handled with the corrected scheme.
The Gauss linear scheme is also used for the divergence appearing in the viscous term (including both molecular and SGS contributions).
Next, we consider the two main cases of numerics used in the campaign, which will serve as a platform for investigating the performance of SGS and wall models for different mesh resolutions.
These cases differ in the discretization of the divergence appearing in the convective term.
The focus on this particular discretization is based on the observation that it has a large influence on the simulation results \citep{mukha2019library}.
The two discretizations are the Gauss linear scheme and the LUST (Linear Upwind Stabilized Transport) scheme \citep{weller2012controlling}.
Therefore, we will refer to these two cases as BG (Base-comb. Gauss) and BL (Base-comb. LUST). 
The LUST scheme combines the Gauss linear and (second-order) upwind approaches, with a weighting of 75\% and 25\%, respectively.
Including upwinding in the LUST scheme adds numerical dissipation, which can help stabilize the solutions.
This is especially relevant when unstructured meshes are considered \citep{martinez2015influence}.
However, one disadvantage of such upwinding approaches in LES is the attenuation of energy at higher wavenumbers, see, e.g., \citep{mittal1997suitability}.
Both cases use the implicit backward scheme for time-stepping (see, e.g., \citep{jasak1996error}) and the preconditioned bi-conjugate gradient (PBiCGStab) \citep{van1992bi, barrett1994templates} for the iterative solution of the momentum and pressure equations.
For the iterative solvers, we use the Diagonal-based Incomplete Cholesky (DIC) preconditioner (pressure equation) and the Diagonal-based Incomplete LU (DILU) preconditioner (momentum equation).
Further, the simulations are carried out with a variable time step that ensures a maximum Courant-Friedrichs-Lewy (CFL) number \citep{courant1967partial} of 0.4.
We note that additional cases that investigate the sensitivity of the BG and BL cases to additional changes in numerics, i.e., temporal discretization, iterative solvers, and the CFL number, were also included in the campaign.
The simulation results were not very sensitive to these changes, as detailed in \ref{app:Add_num_tests}.

\begingroup
\setlength{\tabcolsep}{6pt} 
\renewcommand{\arraystretch}{1.2} 
\begin{table}[ht]
\centering
\begin{tabular}{|ccccc|}
\hline
\multicolumn{5}{|c|}{\textbf{General numerics}}                                                                                                           \\ \hline
\multicolumn{1}{|c|}{\textbf{solver}} &
  \multicolumn{1}{c|}{\textbf{gradient}} &
  \multicolumn{1}{c|}{\textbf{Laplacian}} &
  \multicolumn{1}{c|}{\textbf{surface-normal gradient}} &
  \textbf{divergence (visc. term)} \\ \hline
\multicolumn{1}{|c|}{pisoFOAM}    & \multicolumn{1}{c|}{Gauss linear} & \multicolumn{1}{c|}{Gauss linear} & \multicolumn{1}{c|}{corrected} & Gauss linear \\ \hline
\multicolumn{5}{|c|}{\textbf{BG and BL cases}}                                                                                                            \\ \hline
\multicolumn{1}{|c|}{} &
  \multicolumn{1}{c|}{\textbf{temporal}} &
  \multicolumn{1}{c|}{\textbf{divergence (conv. term)}} &
  \multicolumn{1}{c|}{\textbf{U eq. \& p eq.}} &
  \textbf{CFL} \\ \hline
\multicolumn{1}{|c|}{\textbf{BG}} & \multicolumn{1}{c|}{backward}     & \multicolumn{1}{c|}{Gauss linear} & \multicolumn{1}{c|}{PBiCGStab} & 0.4          \\ \hline
\multicolumn{1}{|c|}{\textbf{BL}} & \multicolumn{1}{c|}{backward}     & \multicolumn{1}{c|}{LUST}         & \multicolumn{1}{c|}{PBiCGStab} & 0.4          \\ \hline
\end{tabular}
\caption{Overview of numerics used in the simulation cases.}
\label{tab:overview_numerics}
\end{table}
\endgroup

Having covered the numerics, we now present an overview of the cases simulated in the campaign.
The details regarding domain and meshes are summarized in Table \ref{tab:mesh_info} for the reader's convenience.
Additional details, e.g., Reynolds numbers and averaging times, are discussed in the text.

\begin{figure}[ht]
    \centering
    \includegraphics[width=1\linewidth]{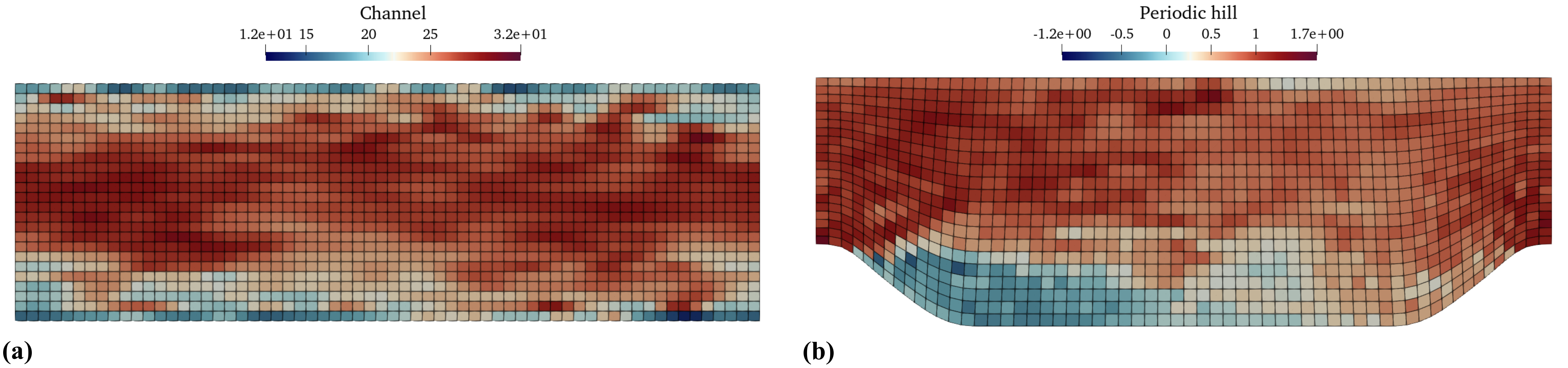}
    \caption{Illustration of the coarse mesh (see Table \ref{tab:mesh_info} for details) for (a) channel flow and (b) periodic hill flow.
    For both cases, the coloring is according to the instantaneous streamwise velocity component.}
    \label{fig:mesh_illustration}
\end{figure}

We perform WMLES of turbulent channel flow at friction Reynolds number $Re_\tau = 5200$, where $Re_\tau = u_\tau \delta / \nu$ with $\delta$ the half-channel width.
This Reynolds number is chosen because it is high enough to be suitable for WMLES and since DNS reference data is readily available \citep{lee2015direct, graham2016web}.
All simulations are performed in a box of size $L_x / \delta \times L_y / \delta \times L_z / \delta = 2\pi \times 2 \times \pi$.
We note that this box size is sufficient for reproducing the one-point statistics of larger boxes up to $Re_\tau = 4200$ \citep{lozano2014effect}, and thus, we expect any errors arising from the domain size to be small.
We use hexahedral meshes with almost isotropic elements (maximum aspect ratio of 1.2), which was found to be optimal in \citep{mukha2019library}.
Two different levels of mesh refinement are considered, which we will refer to as the coarse (C) mesh, $N_x \times N_y \times N_z = 64 \times 24 \times 32$, and the fine (F) mesh, $N_x \times N_y \times N_z = 128 \times 48 \times 64$, respectively.
For all channel flow simulations, the flow is driven by a constant imposed pressure gradient, which balances the wall shear stress according to $\tau_w / \delta = |\partial_x p|$ \citep{Pope_2000}.
An illustration of the coarse mesh for the channel cases is shown in Figure \ref{fig:mesh_illustration} (a).
Similarly, we perform WMLES of the periodic hill case using the same Reynolds number, hill geometry, and domain size as in \citep{frohlich2005highly}.
Specifically, the bulk Reynolds number is $Re_H = 10595$, where $Re_H = U_b H / \nu$ with $U_b$ the bulk velocity at the crest of the hill and $H$ the hill height.
The bounding box for the simulations is $L_x/H \times L_y/H \times L_z/H = 9 \times 3.035 \times 4.5$.
The mesh is structured with hexahedral elements and contains $N_x \times N_y \times N_z = 60 \times 20 \times 30$ elements along each coordinate direction for the coarse (C) mesh and  $N_x \times N_y \times N_z = 120 \times 40 \times 60$ elements for the fine (F) mesh.
Like the channel flow cases, the mesh is designed to minimize the anisotropy of the elements.
Still, the presence of the hill makes some anisotropy unavoidable. 
For both the coarse and fine meshes, the maximum aspect ratio across all elements is about 2.3.
Similarly, the non-orthogonality of the mesh is also quite low.
We also note that for all periodic hill simulations, the flow is driven by a time-dependent forcing term, which takes the same value over the entire domain.
The forcing is adjusted on the fly to ensure a given block velocity $U_b$ over the crest of the hill.
An illustration of the coarse mesh for the periodic hill cases is shown in Figure \ref{fig:mesh_illustration} (b).
We also comment on the averaging times used for both channel and periodic hill cases.
These are chosen to be large enough to ensure that errors from a lack of statistical convergence are minimized.
In terms of flow through times $T_f$, we use averaging times of $T_f = 50$ and $T_f = 200$ for the channel and periodic hill cases, respectively.

\begingroup
\setlength{\tabcolsep}{6pt} 
\renewcommand{\arraystretch}{1.2} 
\begin{table}[ht]
\centering
\begin{tabular}{|cccc|ccc|}
\hline
\multicolumn{4}{|c|}{\textbf{Channel flow}} &
  \multicolumn{3}{c|}{\textbf{Periodic hill flow}} \\ \hline
\multicolumn{1}{|c|}{\textbf{Cases}} &
  \multicolumn{1}{c|}{\textbf{Domain}} &
  \multicolumn{1}{c|}{\textbf{Mesh}} &
  \textbf{SGS} &
  \multicolumn{1}{c|}{\textbf{Domain}} &
  \multicolumn{1}{c|}{\textbf{Mesh}} &
  \textbf{SGS} \\ \hline
\multicolumn{1}{|c|}{\textbf{BG-C-W}} &
  \multicolumn{1}{c|}{$2\pi \times 2 \times \pi$} &
  \multicolumn{1}{c|}{$64 \times 24 \times 32$} &
  WALE &
  \multicolumn{1}{c|}{$9 \times 3.035 \times 4.5$} &
  \multicolumn{1}{c|}{$60 \times 20 \times 30$} &
  WALE \\ \hline
\multicolumn{1}{|c|}{\textbf{BL-C-W}} &
  \multicolumn{1}{c|}{$2\pi \times 2 \times \pi$} &
  \multicolumn{1}{c|}{$64 \times 24 \times 32$} &
  WALE &
  \multicolumn{1}{c|}{$9 \times 3.035 \times 4.5$} &
  \multicolumn{1}{c|}{$60 \times 20 \times 30$} &
  WALE \\ \hline
\multicolumn{1}{|c|}{\textbf{BG-F-W}} &
  \multicolumn{1}{c|}{$2\pi \times 2 \times \pi$} &
  \multicolumn{1}{c|}{$128 \times 48 \times 64$} &
  WALE &
  \multicolumn{1}{c|}{$9 \times 3.035 \times 4.5$} &
  \multicolumn{1}{c|}{$120 \times 40 \times 60$} &
  WALE \\ \hline
\multicolumn{1}{|c|}{\textbf{BL-F-W}} &
  \multicolumn{1}{c|}{$2\pi \times 2 \times \pi$} &
  \multicolumn{1}{c|}{$128 \times 48 \times 64$} &
  WALE &
  \multicolumn{1}{c|}{$9 \times 3.035 \times 4.5$} &
  \multicolumn{1}{c|}{$120 \times 40 \times 60$} &
  WALE \\ \hline
\multicolumn{1}{|c|}{\textbf{BG-C-DS}} &
  \multicolumn{1}{c|}{$2\pi \times 2 \times \pi$} &
  \multicolumn{1}{c|}{$64 \times 24 \times 32$} &
  Dyn. Smag. &
  \multicolumn{1}{c|}{$9 \times 3.035 \times 4.5$} &
  \multicolumn{1}{c|}{$60 \times 20 \times 30$} &
  Dyn. Smag. \\ \hline
\multicolumn{1}{|c|}{\textbf{BL-C-DS}} &
  \multicolumn{1}{c|}{$2\pi \times 2 \times \pi$} &
  \multicolumn{1}{c|}{$64 \times 24 \times 32$} &
  Dyn. Smag. &
  \multicolumn{1}{c|}{$9 \times 3.035 \times 4.5$} &
  \multicolumn{1}{c|}{$60 \times 20 \times 30$} &
  Dyn. Smag. \\ \hline
\multicolumn{1}{|c|}{\textbf{BG-F-DS}} &
  \multicolumn{1}{c|}{$2\pi \times 2 \times \pi$} &
  \multicolumn{1}{c|}{$128 \times 48 \times 64$} &
  Dyn. Smag. &
  \multicolumn{1}{c|}{$9 \times 3.035 \times 4.5$} &
  \multicolumn{1}{c|}{$120 \times 40 \times 60$} &
  Dyn. Smag. \\ \hline
\multicolumn{1}{|c|}{\textbf{BL-F-DS}} &
  \multicolumn{1}{c|}{$2\pi \times 2 \times \pi$} &
  \multicolumn{1}{c|}{$128 \times 48 \times 64$} &
  Dyn. Smag. &
  \multicolumn{1}{c|}{$9 \times 3.035 \times 4.5$} &
  \multicolumn{1}{c|}{$120 \times 40 \times 60$} &
  Dyn. Smag. \\ \hline
\end{tabular}
\caption{Overview of the simulation cases included in the campaign.
The values in the "Domain" column is normalized the the half-channel height $\delta$ (channel) and the hill height $H$ (periodic hill), respectively.}
\label{tab:mesh_info}
\end{table}
\endgroup

Finally, we discuss the relevant SGS and wall modeling details.
Starting with SGS models, an overview of which SGS models are used in which simulation cases is given in Table \ref{tab:mesh_info}.
For both WALE and DS models, the filtering length scale for each element is calculated as $\Delta = V_e^{1/3}$, where $V_e$ is the volume of the element.
Further, the test filter in the DS model
is a simple top-hat filter implemented as a surface integral over the element faces using the face-interpolated values.
Similarly, the DS coefficient averaging is computed as a local average of interpolated face values to prevent numerical instabilities.
Regarding wall modeling, we first provide an overview of the four wall models used in the simulation campaign, which are summarized in Table \ref{tab:overview_wall_models}.
The first wall model (WM1) uses the standard van Driest eddy-viscosity model from Eq. \eqref{eq:van_Driest_eddy_viscosity} and includes no non-equilibrium effects, i.e., $F_i = 0$ in Eq. \eqref{eq:tauw_formula}.
It will, therefore, serve as the EWM baseline.
The second wall model (WM2) uses the pressure-augmented Duprat eddy-viscosity model from Eq. \eqref{eq:Duprat_eddy_viscosity} and includes no non-equilibrium effects, i.e., $F_i = 0$ in Eq. \eqref{eq:tauw_formula}.
The third wall model (WM3) uses the van Driest eddy-viscosity from Eq. \eqref{eq:van_Driest_eddy_viscosity} but includes pressure gradient effects, i.e., $F_i = \partial_i \langle p \rangle$ in \eqref{eq:tauw_formula}.
The fourth wall model (WM4) uses the Duprat eddy-viscosity from Eq. \eqref{eq:Duprat_eddy_viscosity} and also includes pressure gradient effects, i.e., $F_i = \partial_i \langle p \rangle$ in \eqref{eq:tauw_formula}.
Further, as discussed in Section \ref{subsec:wall_models}, a closed-form expression for the wall shear stress can be derived for all four ODE wall models considered, i.e., Eq. \eqref{eq:tauw_formula}.
When this equation is solved numerically using an iterative approach, several different parameters must be specified.
This includes the number of grid points used to evaluate the integral, a reasonable tolerance to use as the stopping condition for the iteration, and the maximum number of iterations.
To establish reasonable values for these parameters, Eq. \eqref{eq:tauw_formula} has been solved in an \textit{a-priori} setting using input data from DNS of both turbulent channel \citep{lee2015direct} and periodic hill \citep{xiao2020flows} flows. 
Based on these tests, we find that a mesh with a single point inside the viscous sub-layer ($y^+ \leq 5$), a tolerance of 0.001, and using 20 iterations is sufficient.

\begingroup
\setlength{\tabcolsep}{6pt} 
\renewcommand{\arraystretch}{1.2} 
\begin{table}[ht]
\centering
\begin{tabular}{|c|c|c|c|c|}
\hline
\textbf{Wall models} &
  WM1 &
  WM2 &
  WM3 &
  WM4 \\ \hline
Eddy-viscosity &
  van Driest: Eq. \eqref{eq:van_Driest_eddy_viscosity} &
  Duprat: Eq. \eqref{eq:Duprat_eddy_viscosity} &
  van Driest: Eq. \eqref{eq:van_Driest_eddy_viscosity} &
  Duprat: Eq. \eqref{eq:Duprat_eddy_viscosity} \\ \hline
\begin{tabular}[c]{@{}c@{}}Non-equilibrium\\ terms\end{tabular} &
  \begin{tabular}[c]{@{}c@{}}Not included:\\ $F_i = 0$ in Eq. \eqref{eq:tauw_formula}\end{tabular} &
  \begin{tabular}[c]{@{}c@{}}Not included:\\ $F_i = 0$ in Eq. \eqref{eq:tauw_formula}\end{tabular} &
  \begin{tabular}[c]{@{}c@{}}Included:\\ $F_i = \partial_i \langle p \rangle$ in Eq. \eqref{eq:tauw_formula}\end{tabular} &
  \begin{tabular}[c]{@{}c@{}}Included:\\ $F_i = \partial_i \langle p \rangle$ in Eq. \eqref{eq:tauw_formula}\end{tabular} \\ \hline
\end{tabular}
\caption{Overview of the four different wall models investigated in the campaign.}
\label{tab:overview_wall_models}
\end{table}
\endgroup

\section{Results}
\label{sec:results}
Results from the planned WMLES campaign are presented and discussed.
The details of the campaign are given in Section \ref{sec:sim_details}, and an overview of the simulation cases is found in Table \ref{tab:mesh_info}.
We first present results from turbulent channel flow in Section \ref{subsec:channel_results} and then present results from turbulent flow over periodic hills in Section \ref{subsec:perhill_results}.
Note that additional cases that investigate the sensitivity of the BG and BL cases to further changes in numerics have also been carried out.
It was found that the simulation results are only weakly sensitive to these additional changes; see \ref{app:Add_num_tests}.

\subsection{A-posteriori results for channel flow}
\label{subsec:channel_results}

We present the results from the different BG and BL cases, see Tables \ref{tab:overview_numerics} and \ref{tab:mesh_info}, using the four different ODE wall models summarized in Table \ref{tab:overview_wall_models}.
The performance is quantified using DNS reference data from the simulation in \citep{lee2015direct}, which is taken as the ground truth.
The data can be accessed through the Johns Hopkins Turbulence Databases \citep{ graham2016web}.
We primarily focus on the cases using the WALE SGS model, BG-C/F-W and BL-C/F-W, while the cases using the DS SGS model, BG-C/F-DS and BL-C/F-DS, are covered briefly at the end of the section.

\begin{figure}[ht]
    \centering
    \includegraphics[width=0.75\textwidth]{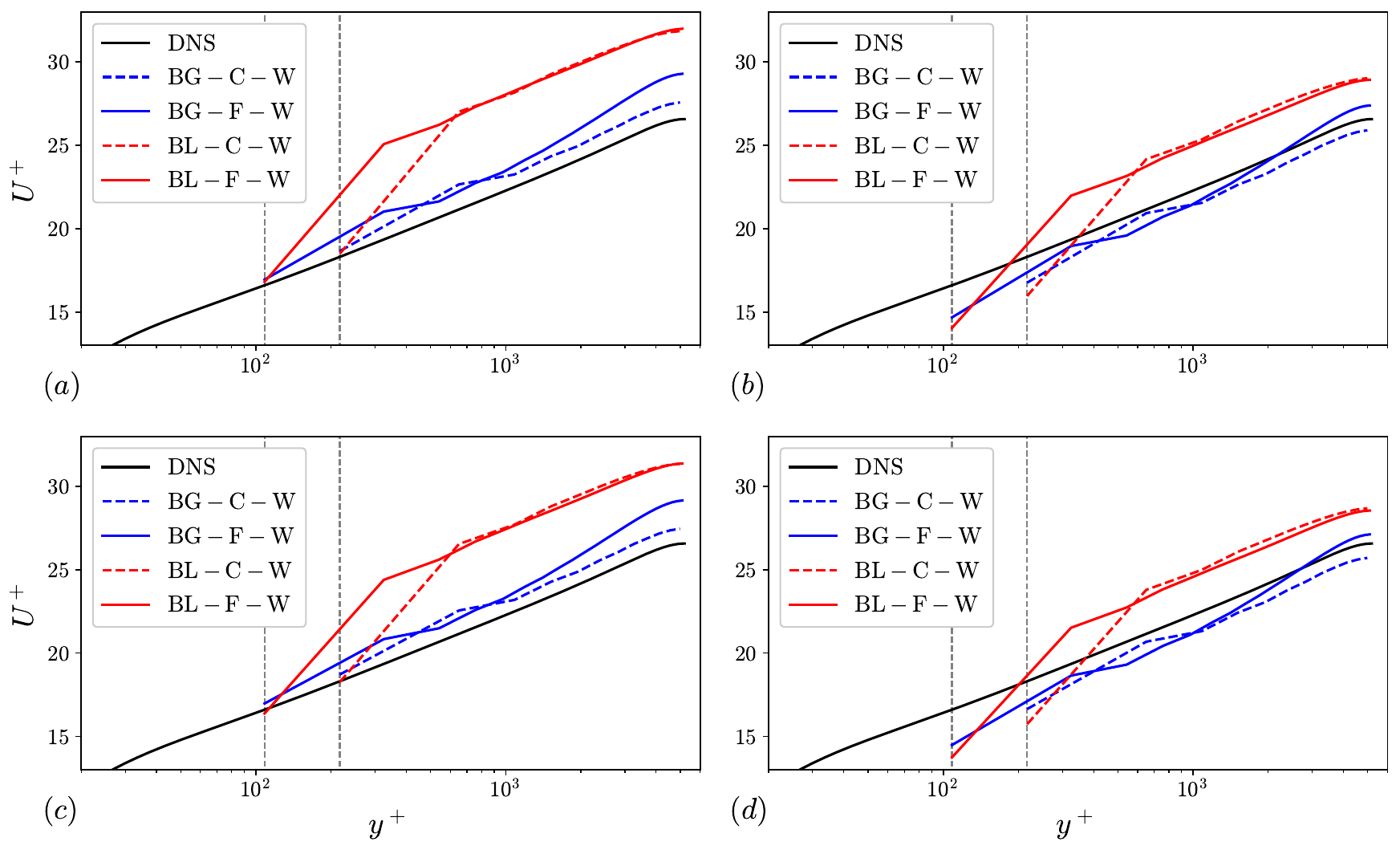}
    \caption{The mean velocity profile from WMLES of turbulent channel flow at $Re_\tau = 5200$.
    The cases are BG-C/F-W and BL-C/F-W, which are summarized in Tables \ref{tab:overview_numerics} and \ref{tab:mesh_info}.
    The wall models are: $(a)$ WM1, $(b)$ WM2, $(c)$ WM3, and $(d)$ WM4, see Table \ref{tab:overview_wall_models}.}
    \label{fig:U_chan_WALE}
\end{figure}

We first consider the mean velocity profile shown in Figure \ref{fig:U_chan_WALE}.
The general observation is that the BG cases show much better agreement with the DNS reference data than the BL cases.
Specifically, the BL cases significantly over-predict the mean velocity profile at all but the first off-wall cell center.
This is consistent with the observations in \citep{mukha2019library} where they forced the channel flow by imposing a constant bulk velocity and used an algebraic EWM.
In this case, when using the LUST scheme, they found that the mean velocity was accurate in the core of the flow but was severely under-predicted at the first off-wall cell center.
If we now compare Figure \ref{fig:U_chan_WALE} (a) and (c), we see only slight deviations between these cases.
This shows that adding the pressure gradient to the ODE wall model has only a small effect on its prediction of the mean velocity in equilibrium flows.
On the other hand, if we now compare Figure \ref{fig:U_chan_WALE} (a) and (b) (or (c) and (d)), it is clear that changing the eddy-viscosity used in the wall model between the van Driest and Duprat models results in noticeable changes in the mean velocity.
By further examination, it becomes evident that the main difference is a vertical shift of the mean velocity profile stemming from the additional eddy-viscosity added to the wall model when using the Duprat model.
Finally, if we compare Figure \ref{fig:U_chan_WALE} (b) and (d), we again observe that the profiles are very similar.
This further confirms the observation that adding the pressure gradient term in the wall model has a negligible effect on the prediction of the mean velocity in equilibrium flows.

\begin{figure}[ht]
    \centering
    \includegraphics[width=0.75\textwidth]{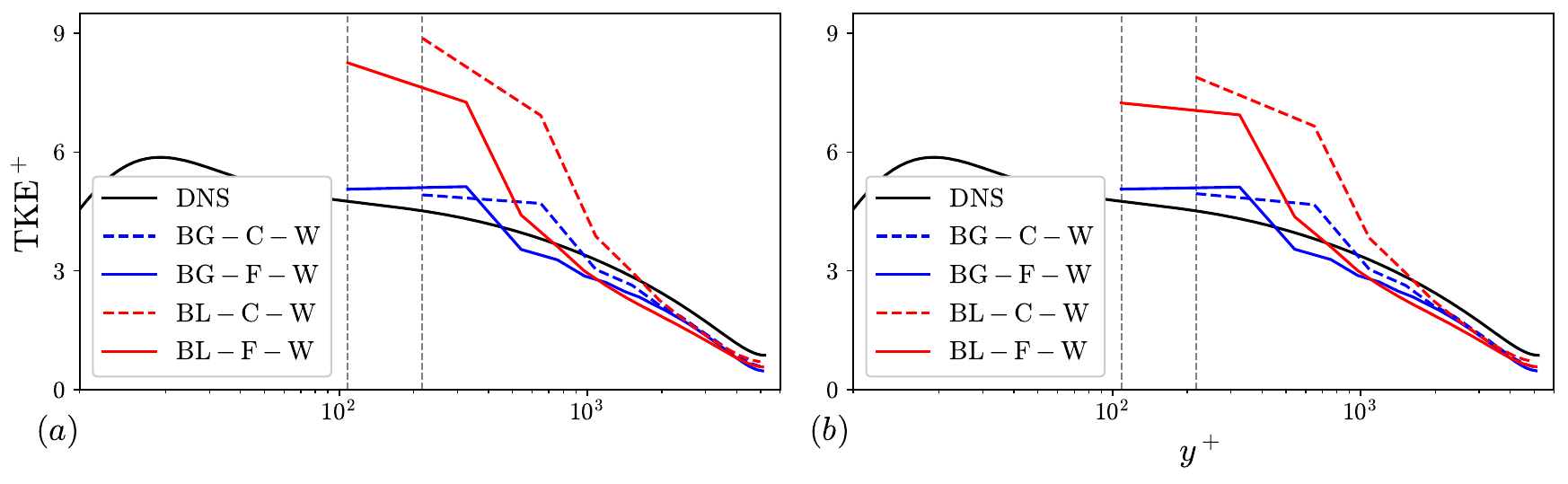}
    \caption{The mean TKE profile from WMLES of turbulent channel flow at $Re_\tau = 5200$.
    The cases are BG-C/F-W and BL-C/F-W, which are summarized in Tables \ref{tab:overview_numerics} and \ref{tab:mesh_info}.
    The wall models are: $(a)$ WM1 and $(b)$ WM4, see Table \ref{tab:overview_wall_models}.}
    \label{fig:TKE_chan_WALE}
\end{figure}

The performance of the BG-C/F-W and BL-C/F-W cases is further quantified by considering the mean turbulent kinetic energy (TKE) predictions as shown in Figure \ref{fig:TKE_chan_WALE}.
Note that we only show the cases using wall models WM1 and WM4 because the results for the remaining cases were very similar.
We see that the BG cases show much better agreement with the DNS reference data than the BL cases.
Specifically, the BL cases significantly over-predict the TKE profile at the first two off-wall cell centers.
This increase in the TKE level for the BL cases is somewhat puzzling, as the additional numerical dissipation inherent to the LUST scheme would be expected to dampen the turbulent fluctuations.
Therefore, the results indicate that there is some mechanism(s) present in the BL cases, which creates a mismatch in the TKE budget in the near-wall region.
This question is addressed further below.
Regarding wall modeling, we see that the two different wall models, WM1 and WM2, give similar TKE results with slight differences in the first couple of near-wall cell centers.
This illustrates that the sensitivity of TKE predictions to wall modeling is small in equilibrium flows.

\begin{figure}[ht]
    \centering
    \includegraphics[width=0.85\textwidth]{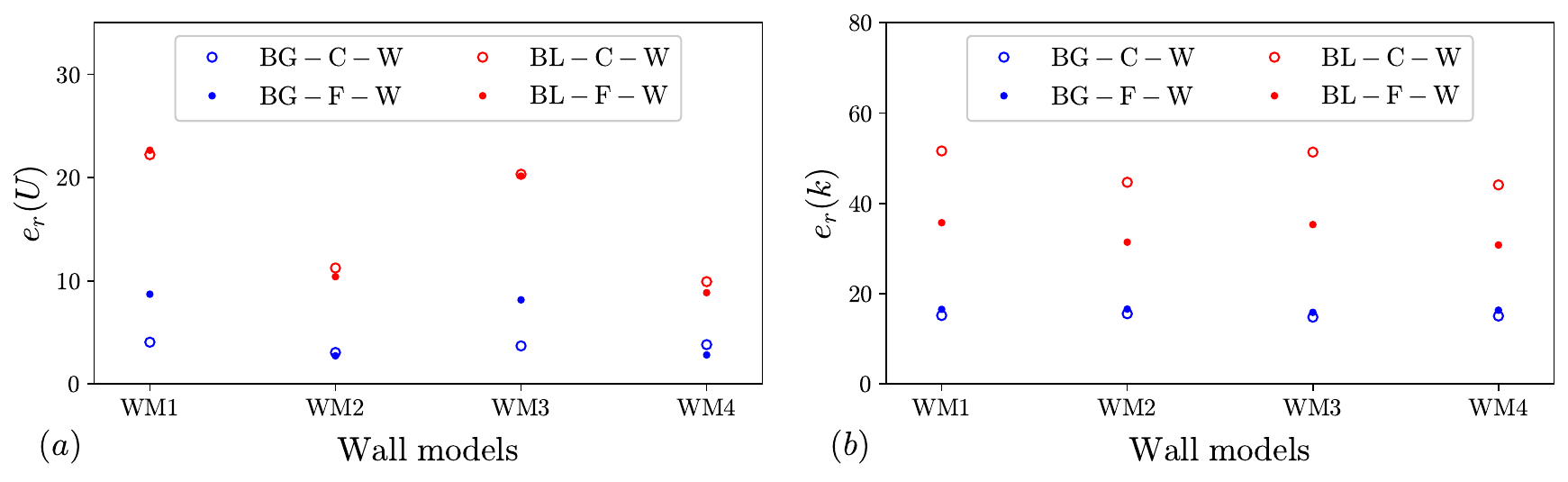}
    \caption{The percentage error for the mean velocity and TKE from WMLES of turbulent channel flow at $Re_\tau = 5200$.
    The cases are BG-C/F-W and BL-C/F-W, which are summarized in Tables \ref{tab:overview_numerics} and \ref{tab:mesh_info}.
    The different wall models are summarized in Table \ref{tab:overview_wall_models}.}
    \label{fig:U_TKE_err_chan_WALE}
\end{figure}

The performance of the BG-C/F-W and BL-C/F-W cases is summarized in terms of mean velocity and TKE errors.
Inspired by \citep{lozano2019error}, we define the errors as
\begin{align}
    e_r(U) = 100 \times \left[ \frac{\int_h^\delta (U - U_\text{ref})^2 \, dy}{\int_h^\delta U_\text{ref}^2 \, dy}  \right]^{1/2} , \qquad e_r(k) = 100 \times \left[ \frac{\int_h^\delta (k - k_\text{ref})^2 \, dy}{\int_h^\delta k_\text{ref}^2 \, dy}  \right]^{1/2} ,
\label{eq:errors_U_TKE}
\end{align}
where $k$ is the TKE.
Note that we include a factor of $100$ to get a percentage error.
The errors are shown in Figure \ref{fig:U_TKE_err_chan_WALE}.
For the mean velocity errors in Figure \ref{fig:U_TKE_err_chan_WALE} (a), 
we see that the BG-C/F-W cases perform consistently better than the BL-C/F-W cases.
Further, we also observe that among the different wall models, the ones using the Duprat eddy-viscosity, WM2 and WM4, are more accurate.
In terms of the TKE shown in Figure \ref{fig:U_TKE_err_chan_WALE} (b), the BG-C/F-W cases also perform better than the BL-C/F-W cases, however, the errors are generally larger than for the mean velocity.
It is also clear that the TKE errors are less sensitive to wall modeling than the mean velocity errors.

\begin{figure}[ht]
    \centering
    \includegraphics[width=0.75\textwidth]{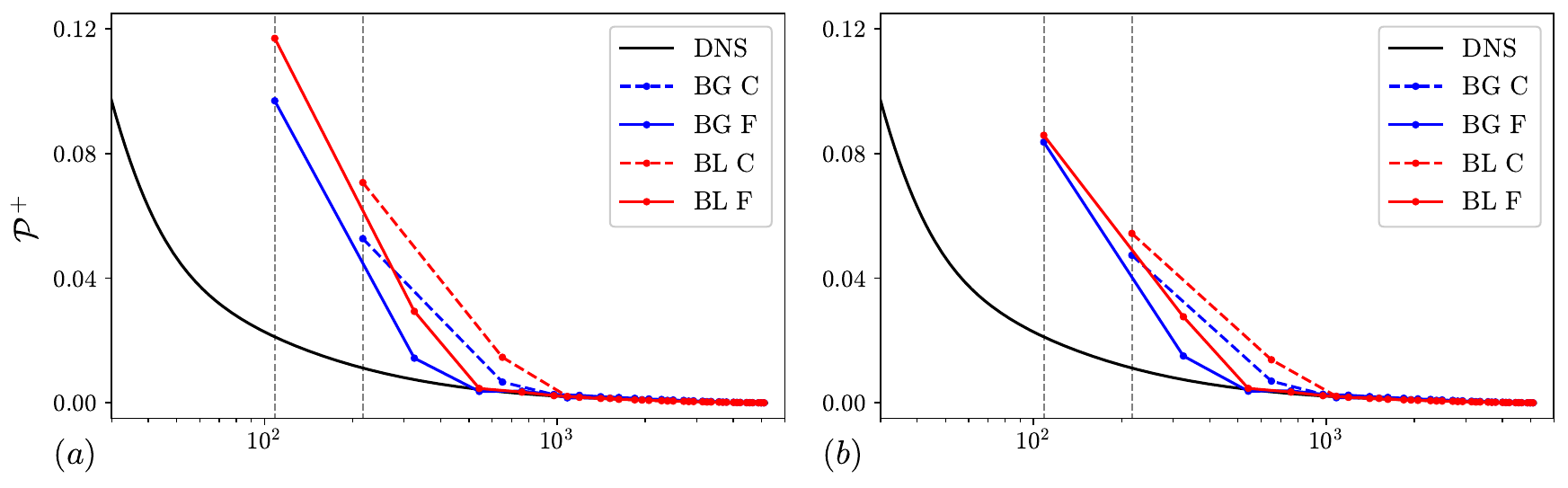}
    \caption{Production of TKE from WMLES of turbulent channel flow at $Re_\tau = 5200$.
    The cases are BG-C/F-W and BL-C/F-W, which are summarized in Tables \ref{tab:overview_numerics} and \ref{tab:mesh_info}.
    The wall models are: $(a)$ WM1 and $(b)$ WM4, see Table \ref{tab:overview_wall_models}.}
    \label{fig:TKE_prod_chan_WALE}
\end{figure}

To investigate the reason for the large discrepancies in the TKE for the BL-C/F-W cases seen in Figure \ref{fig:TKE_chan_WALE}, we first investigate the production of TKE in the near-wall region.
We note that, for turbulent channel flow, the mean production of TKE is given by (see, e.g., \citep{Pope_2000})
\begin{align}
    \mathcal{P} = - \langle \widetilde{u'_1} \widetilde{u'_2} \rangle \partial_2 \langle \tilde{u}_1 \rangle ,
\end{align}
which is proportional to the mean wall-normal gradient.
Looking at Figure \ref{fig:U_chan_WALE}, we see that the mean wall-normal gradient is larger in the first two off-wall cell centers for the BL cases than for the BG cases.
This could, therefore, act as an amplification factor in the production of TKE in the near-wall region.
We examine this empirically for the BG-C/F-W and BL-C/F-W cases using wall models WM1 and WM4.
Figure \ref{fig:TKE_prod_chan_WALE} shows that the production of TKE in the near-wall region for the BL cases is indeed slightly amplified compared to the BG cases.
However, the magnitude of the amplification is quite small compared to the one observed for the TKE from the BL cases in Figure \ref{fig:TKE_chan_WALE}.

\begin{figure}[ht]
    \centering
    \includegraphics[width=0.75\textwidth]{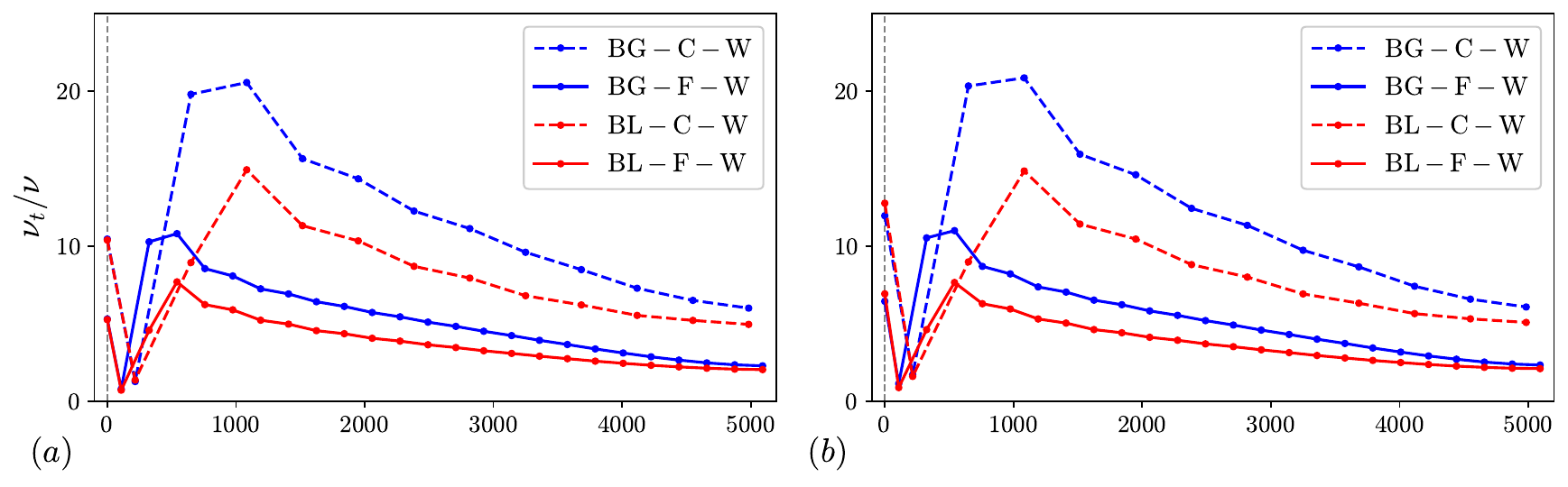}
    \caption{The mean eddy-viscosity from WMLES of turbulent channel flow at $Re_\tau = 5200$.
    The cases are BG-C/F-W and BL-C/F-W, which are summarized in Tables \ref{tab:overview_numerics} and \ref{tab:mesh_info}.
    The wall models are: $(a)$ WM1 and $(b)$ WM4, see Table \ref{tab:overview_wall_models}.}
    \label{fig:nut_chan_WALE}
\end{figure}

Next, we consider the activity of the WALE SGS model.
This is done by looking at the mean turbulent eddy-viscosity, which we use as a proxy for the mean SGS dissipation.
To see why this is a reasonable choice, we note that the mean SGS dissipation is given by \citep{meneveau2000scale}
\begin{align}
    \epsilon^{\text{SGS}} = -\langle \tau_{ij}^{\Delta} \tilde{S}_{ij} \rangle ,
\end{align}
which, for an eddy-viscosity model as in Eq. \eqref{eq:eddy_viscosity_model}, gives
\begin{align}
     \epsilon^{\text{SGS,EV}} = 2\langle \nu_t \tilde{S}_{ij} \tilde{S}_{ij} \rangle .
\end{align}
If we then introduce the Reynolds decompositions $\nu_t = \langle \nu_t \rangle + \nu'_t$ and $\tilde{S}_{ij} \tilde{S}_{ij} = \langle \tilde{S}_{ij} \tilde{S}_{ij} \rangle + {(\tilde{S}_{ij} \tilde{S}_{ij})}'$, the above expression can be rewritten as
\begin{align}
     \epsilon^{\text{SGS,EV}} = 2\langle \nu_t \rangle \langle \tilde{S}_{ij} \tilde{S}_{ij} \rangle + 2\langle \nu'_t {(\tilde{S}_{ij} \tilde{S}_{ij})}' \rangle .
\end{align}
The first term represents the majority of the contribution to SGS dissipation, which shows that mean eddy-viscosity is indeed a reasonable proxy. 
Similar to the TKE, wall modeling was found to have a weak influence on the mean eddy-viscosity, and thus, we only focus on the results from WM1 and WM4.
The mean eddy-viscosity for the BG-C/F-W and BL-C/F-W cases is shown in Figure \ref{fig:nut_chan_WALE}. 
We see that the mean eddy-viscosity is larger for the BG cases than the BL cases, especially in the second and third off-wall cell centers.
This suggests that the WALE model introduces much less SGS dissipation in the near-wall region for the BL cases than for the BG cases.

To conclude the above discussion about the TKE overshoot, the BL cases have a slightly amplified production and a lot less SGS dissipation of TKE in the near-wall region when compared with the BG cases.
We believe that the combination of these two factors explains the overshoot of TKE in the near-wall region observed for the BL cases.
However, it does not explain the small overshoot at the second cell center for the BG cases, see Figure \ref{fig:TKE_chan_WALE}.

\begin{figure}[ht]
    \centering
    \includegraphics[width=0.75\textwidth]{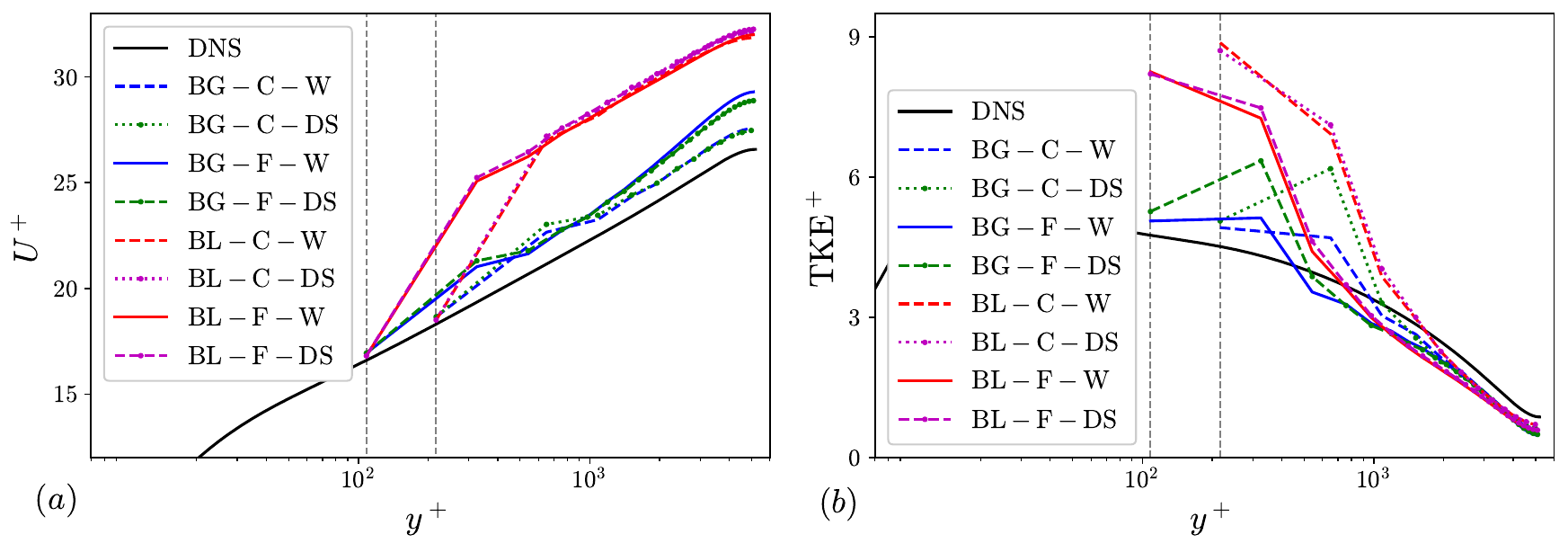}
    \caption{Results from WMLES of turbulent channel flow at $Re_\tau = 5200$: (a) mean velocity and (b) mean TKE.
    The cases are BG-C/F-W/DS and BL-C/F-W/DS, which are summarized in Tables \ref{tab:overview_numerics} and \ref{tab:mesh_info}.
    The wall model used is the EWM, i.e., WM1 in Table \ref{tab:overview_wall_models}.}
    \label{fig:U_chan_SGSTests}
\end{figure}

Finally, we consider the sensitivity of the above results from the BG-C/F-W and BG-C/F-W cases to a change in the SGS model.
For convinience of the presentation, we only consider results from the BG-C/F-DS and BL-C/F-DS cases using the EWM, i.e., WM1 in Table \ref{tab:overview_wall_models}.
From the results in Figure \ref{fig:U_chan_SGSTests} (a), we see that the differences are small for the mean velocity.
From Figure \ref{fig:U_chan_SGSTests} (b), we see that the differences in TKE are small for the BL cases.
This makes sense as the amount of numerical dissipation introduced by the LUST scheme can overwhelm the contribution from the SGS model, as was also observed in \citep{martinez2015influence}.
For the BG cases, however, using the DS SGS model results in a significantly higher TKE value at the second off-wall cell center than using the WALE SGS model.
We do not currently have a satisfactory explanation of this observed overshoot.

\subsection{Flow over periodic hills}
\label{subsec:perhill_results}

We now move on to consider turbulent flow over periodic hills.
First, we present the result from an \textit{a-priori} investigation of wall model performance in Section \ref{subsubsec:a_priori_perhill}.
The investigation uses mean velocity and pressure gradient data from a WRLES of the periodic hill case carried out by the authors.
The WRLES details and validation against the results from \citep{frohlich2005highly} is given in \ref{app:WRLES_results}.
An important part of this investigation is to assess the potential error in neglecting the convective term in the wall models.
After this, we present the \textit{a-posteriori} results from our WMLES simulation campaign in Section \ref{subsubsec:a_posteriori_perhill}.

\subsubsection{A-priori results for periodic hill}
\label{subsubsec:a_priori_perhill}

\begin{figure}[ht]
    \centering
    \includegraphics[width=0.75\textwidth]{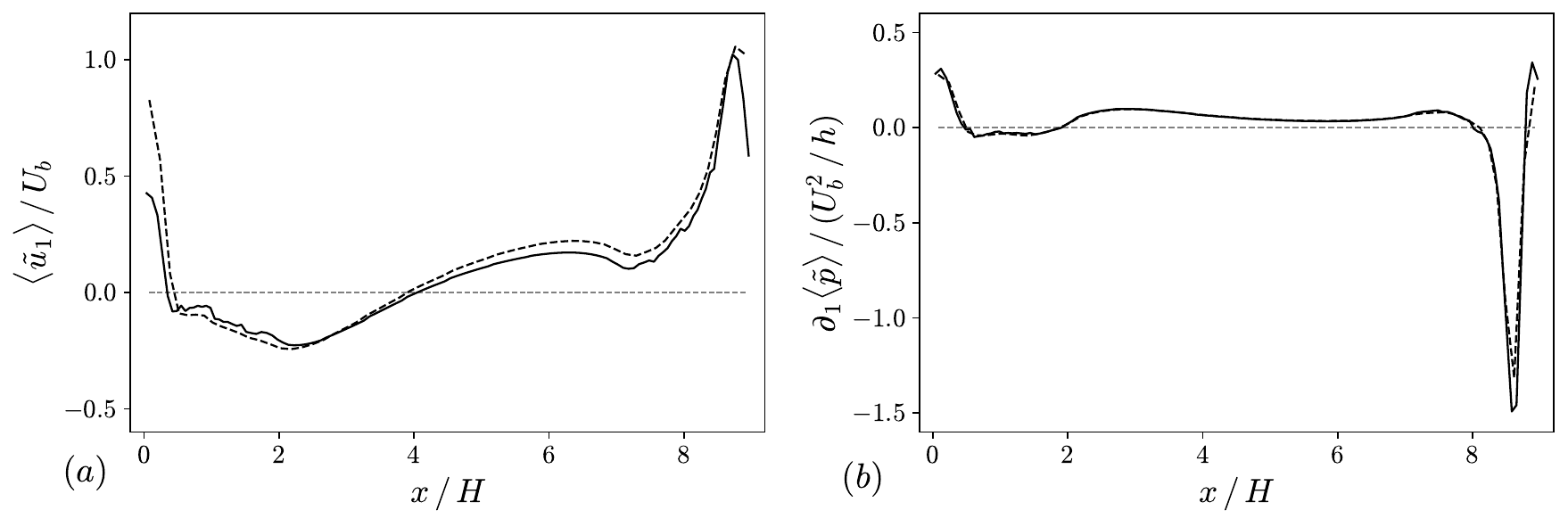}
    \caption{Wall model input for \textit{a-priori} analysis of the periodic hill case.
    The data comes from a WRLES, which is discussed in \ref{app:WRLES_results}.
    Dashed and solid lines correspond to sampling from the coarse (C) and fine (F) WMLES meshes, respectively, see Table \ref{tab:mesh_info}.}
    \label{fig:WM_input_data}
\end{figure}

The input needed to evaluate the ODE wall models discussed in Section \ref{subsec:wall_models}, and summarized in Table \ref{tab:overview_wall_models}, is the local streamwise velocity and pressure gradient components.
Here, we use mean values (averaged in time and along the spanwise direction) taken from the WRLES results discussed in \ref{app:WRLES_results}.
The values are then sampled at locations matching the first off-wall cell centers on the coarse (C) and fine (F) WMLES meshes, see Table \ref{tab:mesh_info}, used in the subsequent \textit{a-posteriori} investigations in Section \ref{subsubsec:a_posteriori_perhill}. 
The input values are shown in Figure \ref{fig:WM_input_data} for completion.
To properly evaluate the performance of the four wall models in Table \ref{tab:overview_wall_models}, we examine both the wall-modeled velocity profiles below the first off-wall cell centers and the wall shear stress predictions.
To obtain the former, direct ODE solvers based on Eq. \eqref{eq:TBLE} have been implemented, with $F_i = 0$ for WM1 and WM2 and $F_i = \partial_i \langle p \rangle$ for WM3 and WM4.
These are solved on a fine near-wall mesh, which has the first off-wall point below $y^+ = 1$, using the input data from Figure \ref{fig:WM_input_data} as the outer boundary condition and with the no-slip condition applied at the wall.
The wall shear stress, on the other hand, is found by iterative solution of Eq. \eqref{eq:tauw_formula} as is done in the actual simulations, again using the input data from Figure \ref{fig:WM_input_data}.

\begin{figure}[ht]
    \centering
    \includegraphics[width=0.75\textwidth]{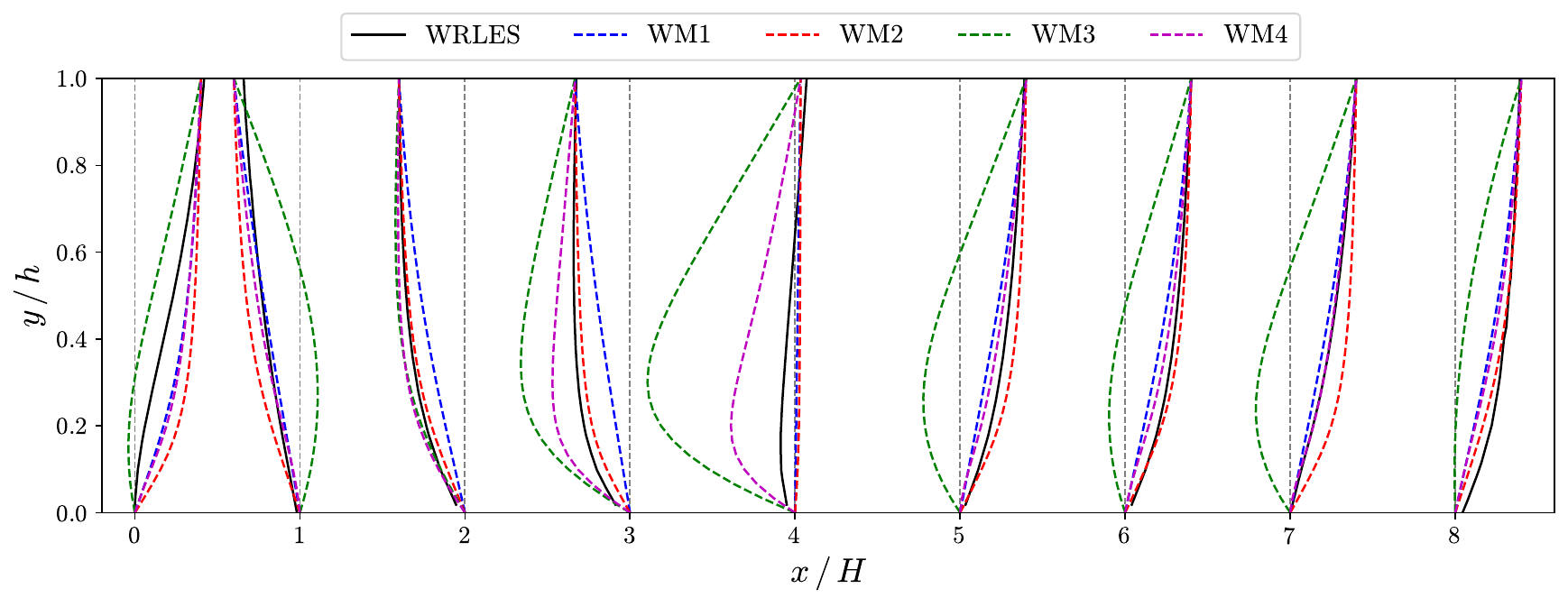}
    \caption{Comparison of \textit{a-priori} wall modeled velocity profiles in the near-wall region from the four different wall models, see Table \ref{tab:overview_wall_models}.
    The reference data is from a WRLES, which is discussed in \ref{app:WRLES_results}.}
    \label{fig:WM_U_results}
\end{figure}

Moving on to the \textit{a-priori} results, we start by considering the local streamwise velocity profiles predicted by each wall model as shown in Figure \ref{fig:WM_U_results}.
The first thing to note is that none of the wall models perform well consistently.
Similarly, it is also noticeable that none of the wall models perform well at $x/H = 0$, as at least one of the models perform well at all the other locations.
Second, we observe that WM3 shows the largest variation and errors and that it significantly overpredicts the length of the separation bubble.
This indicates that wall models using the van Driest eddy-viscosity model in Eq. \eqref{eq:van_Driest_eddy_viscosity} are overly sensitive to inclusion of the pressure gradient.
For the remaining models, we observe that WM1 and WM2 generally perform the best, while WM4 is somewhere in the middle.

\begin{figure}[ht]
    \centering
    \includegraphics[width=0.75\textwidth]{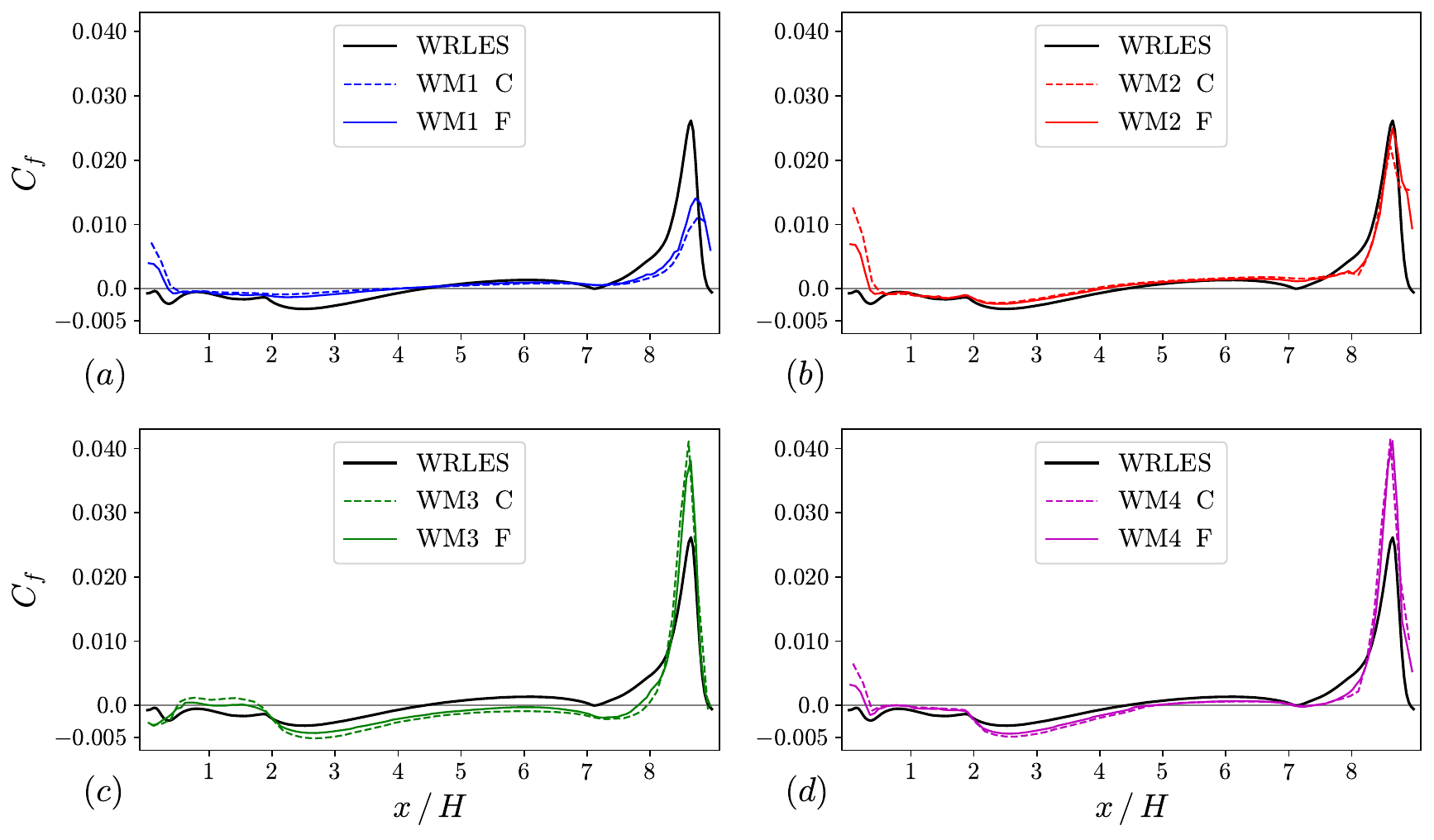}
    \caption{Comparison of \textit{a-priori} predictions of skin friction coefficient $C_f$ from the four different wall models, see Table \ref{tab:overview_wall_models}.
    Dashed and solid lines correspond to results using the coarse (C) and fine (F) input data, respectively, see Figure \ref{fig:WM_input_data}.
    The reference data is from a WRLES, which is discussed in \ref{app:WRLES_results}.}
    \label{fig:WM_tauw_results}
\end{figure}

Next, we consider the wall shear stress predictions.
These are presented in terms of the skin friction coefficient $C_f$, which is defined as
\begin{align}
  C_f(x) = \frac{\tau_w(x)}{\tfrac{1}{2} \rho U_b} .
\label{eq:skin_friction_coeff}
\end{align}
Note that it is implicitly understood that averaging has been performed in time and along the spanwise direction.
The $C_f$ results are shown in Figure \ref{fig:WM_tauw_results}.
From Figure \ref{fig:WM_tauw_results} (a), we see that the EWM, i.e., WM1, hardly predicts separation at all and that it severely underpredicts the maximum $C_f$ value.
On the other hand, Figure \ref{fig:WM_tauw_results} (b) shows that WM2 gives good overall results in terms of both the length of the separation bubble and the maximum $C_f$ value.
We note that the only difference between WM1 and WM2 is the eddy-viscosity models, van Driest Eq. \eqref{eq:van_Driest_eddy_viscosity} and Duprat Eq. \eqref{eq:Duprat_eddy_viscosity}, respectively. 
The remaining results from wall models WM3 and WM4, which both include the pressure gradient term, are shown in Figures \ref{fig:WM_tauw_results} (c) and (d), respectively.
It is observed that both models successfully predict separation but that they strongly overpredict the maximum $C_f$ value.
Of these two models, WM3 performs the worst, because of its hugely enlarged separated region.

Considering the \textit{a-priori} results for both the velocity profiles and wall shear stress presented above, we see that WM2 has the best overall performance of the four different wall models.
For models WM3 and WM4, we observe that inclusion of the pressure gradient term in Eq. \eqref{eq:TBLE} does improve the prediction of separation compared to the EWM, i.e., WM1.
However, it also leads to a severe over-prediction of the maximum wall shear stress.
This illustrates that inclusion of the pressure gradient without any of the other non-equilibrium terms gives mixed results, which is consistent with what has been reported in the literature, see \citep{larsson2016large} and references therein.

\begin{figure}[ht]
    \centering
    \includegraphics[width=0.8\textwidth]{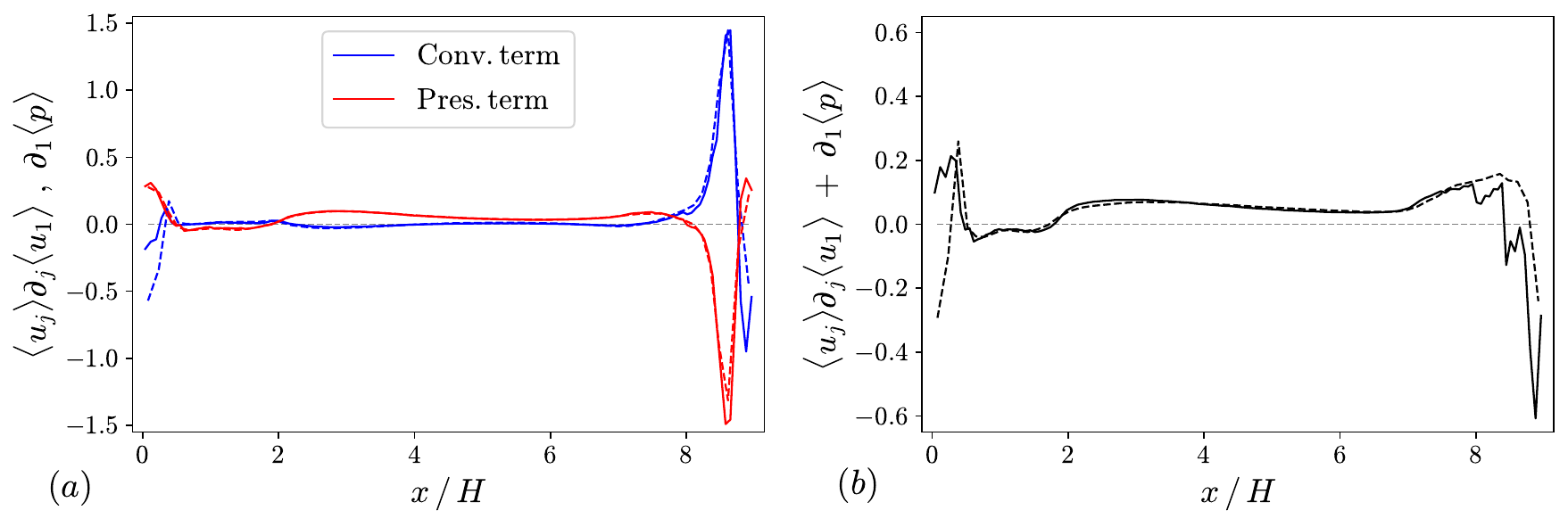}
    \caption{Comparison of (a) the convective and pressure gradient terms and (b) their sum.
    Dashed and solid lines correspond to sampling from the coarse (C) and fine (F) WMLES meshes, respectively, see Table \ref{tab:mesh_info}.
    The data is from a WRLES, which is discussed in \ref{app:WRLES_results}.}
    \label{fig:u_conv_vs_p_grad}
\end{figure}
 
\begin{figure}[ht]
    \centering
    \includegraphics[width=0.98\textwidth]{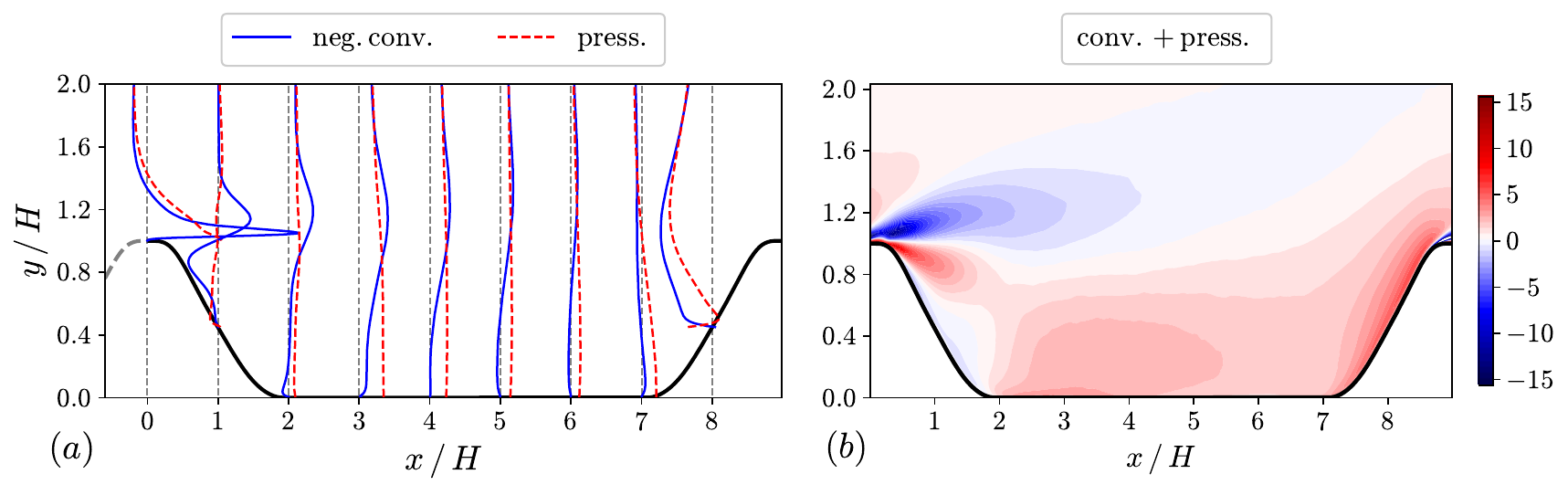}
    \caption{Illustration of (a) near-wall profiles of the convective (solid) and pressure gradient (dashed) terms (the sign of the former has been flipped to ease comparison) and (b) contours of the sum of the two terms (with no flipping of the signs).
    The profiles in (a) are normalized by a factor of 10 to enhance visibility.
    The data is from a WRLES, which is discussed in \ref{app:WRLES_results}.}
    \label{fig:u_conv_vs_p_grad_near_wall}
\end{figure}

We now return to the question of the balance between the convective and pressure gradient terms discussed earlier at the end of Section \ref{subsec:wall_models}.
To investigate this balance for the periodic hill cases, we extract the convective and pressure gradient terms from the WRLES presented in \ref{app:WRLES_results}.
We first consider these terms extracted at positions corresponding to the matching locations used in the subsequent WMLES presented in Section \ref{subsubsec:a_posteriori_perhill}.
These results are shown in Figure \ref{fig:u_conv_vs_p_grad}.
Looking at Figure \ref{fig:u_conv_vs_p_grad} (a), we see that the convective and pressure gradient terms show significant cancellation, especially at the front side of the hill, where both terms have large spikes in their values.
However, if we examine the sum of the convective and pressure gradient terms, which is shown in Figure \ref{fig:u_conv_vs_p_grad} (b), it is clear that the cancellation is not perfect.
This is most evident around the top of the hill, but it is also visible inside the latter half of the separation bubble and in the subsequent recovery region.
To further investigate the degree of balance between the convective and pressure gradient terms, we consider the profiles of each term at several streamwise positions along the flow.
The profiles are extracted along the wall-normal direction in the global coordinate system.
We also consider a contour plot of the sum of the two terms to get a better qualitative understanding of the balance between the two.
From the profiles shown in Figure \ref{fig:u_conv_vs_p_grad_near_wall} (a), we see that the degree to which the terms balance or not varies with the distance to the wall.
This is consistent with previous observations \citep{hickel2013parametrized,larsson2016large,kamogawa2023ordinary} where the balance was observed to hold in the outer part of the boundary layer but break down as the wall is approached.
Physically, this occurs because the convective term, unlike the pressure gradient, satisfies the no-slip condition at the wall.
From the contour plot of the sum of the two terms in Figure \ref{fig:u_conv_vs_p_grad_near_wall} (b),
we make the following observations.
The first is that the terms show a strong imbalance around the hill, especially associated with the shear layer over the backside of the hill.
Second, some degree of imbalance is present in the near-wall region throughout almost the entire domain.
This is largely expected, given that the flow is driven by a pressure gradient.
This persistent imbalance is likely the reason for the poor wall shear stress predictions from the EWM, i.e., WM1, in Figure \ref{fig:WM_tauw_results} (a).
Similarly, the observed imbalance also helps explain why inclusion of the pressure gradient without also including the convective term leads to mixed results, see Figures \ref{fig:WM_tauw_results} (c) and (d).
This suggests that for non-equilibrium wall models to outperform EWMs consistently, they need to include both convective and pressure gradient terms.

\subsubsection{A-posteriori results for periodic hill}
\label{subsubsec:a_posteriori_perhill}

\begin{figure}[ht]
    \centering
    \includegraphics[width=0.95\textwidth]{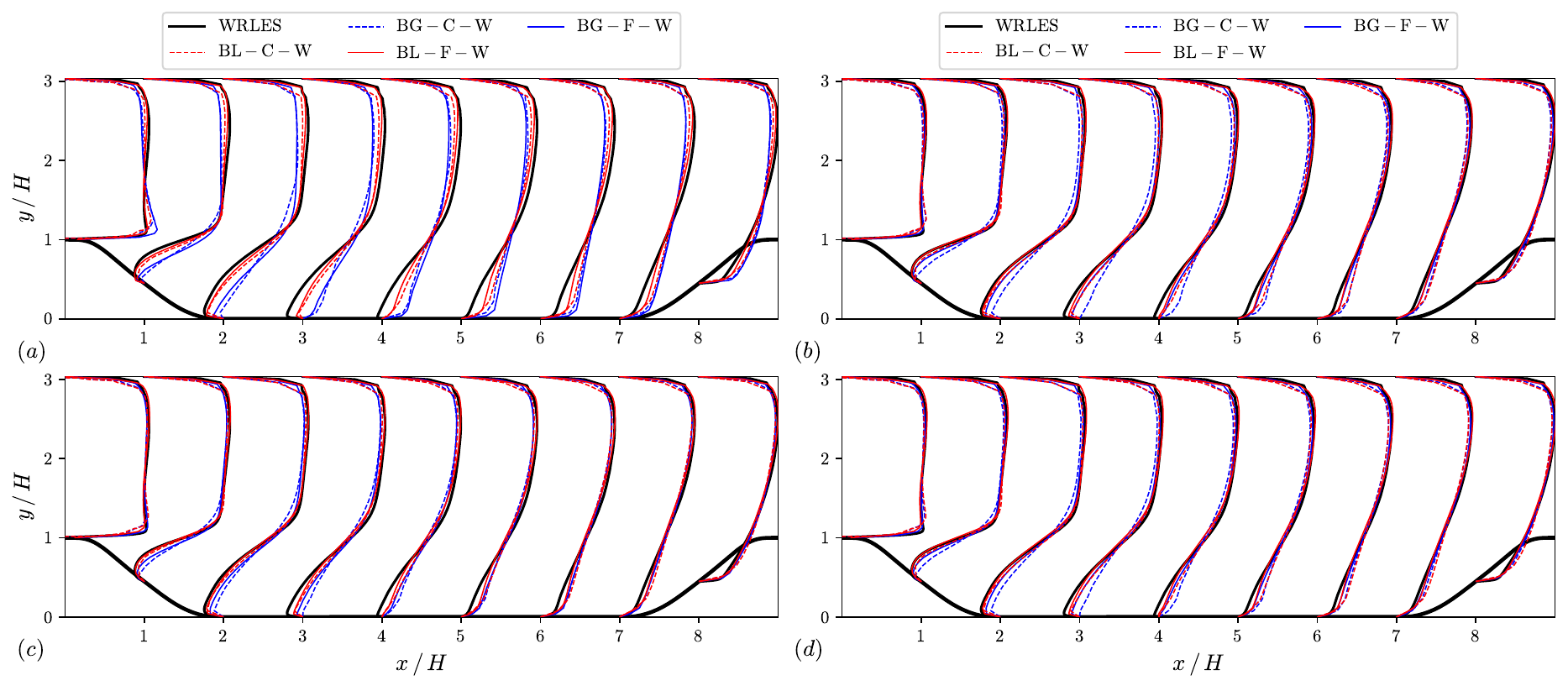}
    \caption{The mean velocity profile from WMLES of periodic hill flow at $Re_H = 10595$.
    The cases are BG-C/F-W and BL-C/F-W, which are summarized in Tables \ref{tab:overview_numerics} and \ref{tab:mesh_info}.
    The wall models are: $(a)$ WM1, $(b)$ WM2, $(c)$ WM3, and $(d)$ WM4, see Table \ref{tab:overview_wall_models}.}
    \label{fig:U_perHill_WALE}
\end{figure}

We now cover the \textit{a-posteriori} results for the periodic hill.
We first consider the BG-C/F-W and BL-C/F-W cases, see Tables \ref{tab:overview_numerics} and \ref{tab:mesh_info}, using the four different wall models summarized in Table \ref{tab:overview_wall_models}. 
Figure \ref{fig:U_perHill_WALE} shows the mean streamwise velocity profiles.
Several general observations can be made.
Firstly, all of the non-equilibrium wall models, i.e., WM2 to WM4, perform better than the EWM (WM1).
Secondly, the mean velocity predictions of the BL-C/F-W cases are better than those of the BG-C/F-W cases.
This is most evident in the separated region and is especially clear for the EWM (WM1) cases in Figure \ref{fig:U_perHill_WALE} (a).
As the only difference between the BL-C/F-W and BG-C/F-W cases is the additional dissipation introduced by the LUST scheme, this observation suggests that the performance of the BG cases could be improved by using a SGS model which is more dissipative in separated regions.
Thirdly, all cases struggle with the latter part of the separation bubble and the subsequent recovery region.
This is more evident for the BG-C/F-W cases than for the BL-C/F-W cases and is especially noticeable when the EWM (WM1) is used.
Finally, we observe that the performance on the fine (F) mesh is generally better than that on the coarse (C) mesh.

\begin{figure}[ht]
    \centering
    \includegraphics[width=0.95\textwidth]{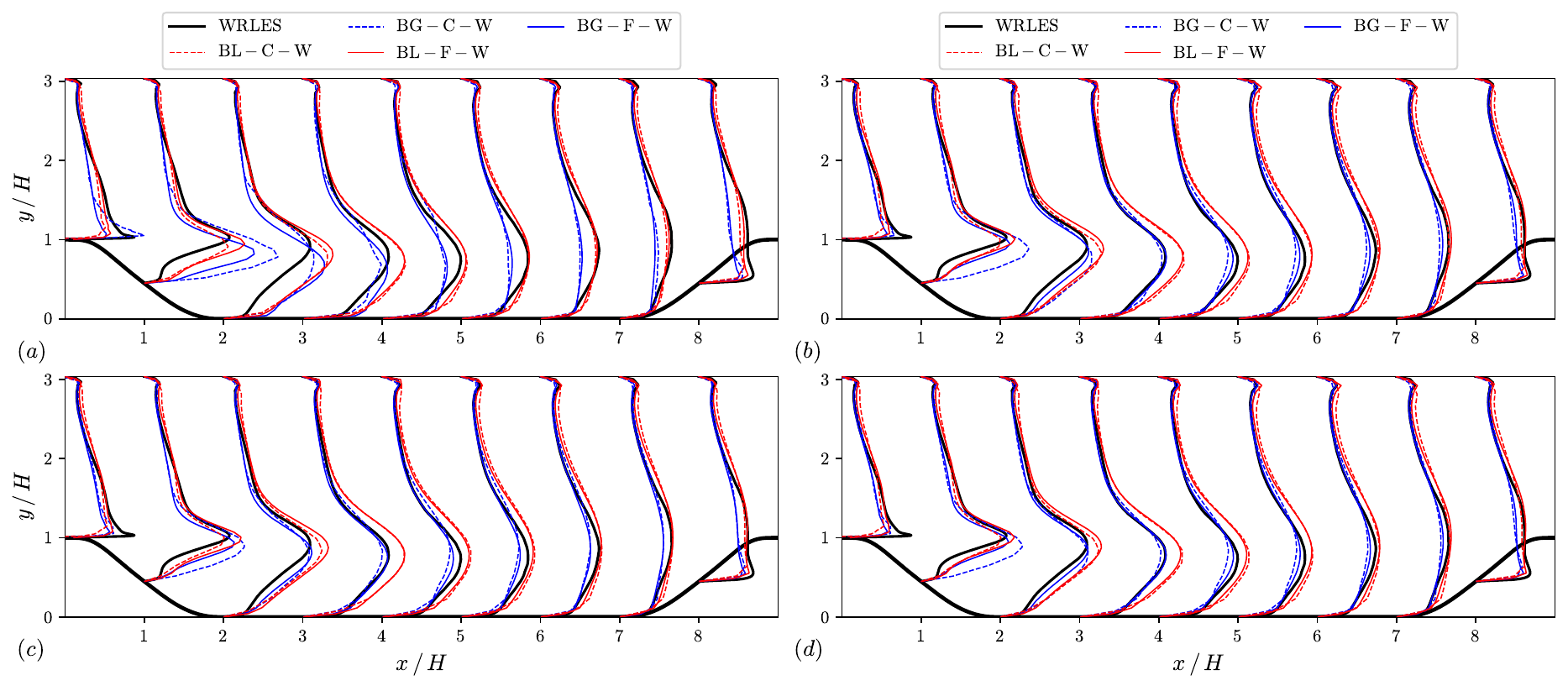}
    \caption{The mean TKE from WMLES of periodic hill flow at $Re_H = 10595$.
    The cases are BG-C/F-W and BL-C/F-W, which are summarized in Tables \ref{tab:overview_numerics} and \ref{tab:mesh_info}.
    The wall models are: $(a)$ WM1, $(b)$ WM2, $(c)$ WM3, and $(d)$ WM4, see Table \ref{tab:overview_wall_models}.}
    \label{fig:TKE_perHill_WALE}
\end{figure}

Next, we investigate the mean TKE, which is shown in Figure \ref{fig:TKE_perHill_WALE}.
Similar to the mean velocity profiles, we see that the largest errors occur in the separated region.
We also see that the EWM (WM1) provides the worst results of the four different wall models.
Interestingly, it is not as evident as for the mean velocity whether the best predictions are from the BG-C/F-W or BL-C/F-W cases.
Specifically, the BL-C/F-W cases seem to give better results when using the EWM (WM1); however, for the remaining wall models, i.e., WM2 to WM4, the BG-C/F-W cases seem to give slightly better overall results.
We also note that similar to the mean TKE result for the channel flow in Figure \ref{fig:TKE_chan_WALE}, the BL-C/F-W cases tend to overestimate the mean TKE, however, not to the same degree.
In terms of meshing, similar to the mean velocity results, we see that the performance improves on the fine (F) mesh compared to the coarse (C) mesh.
This is especially noticeable at $x / H = 1$ where, for all wall models but the EWM (WM1), the TKE is captured much more accurately on the fine mesh.

\begin{figure}[ht]
    \centering
    \includegraphics[width=0.85\textwidth]{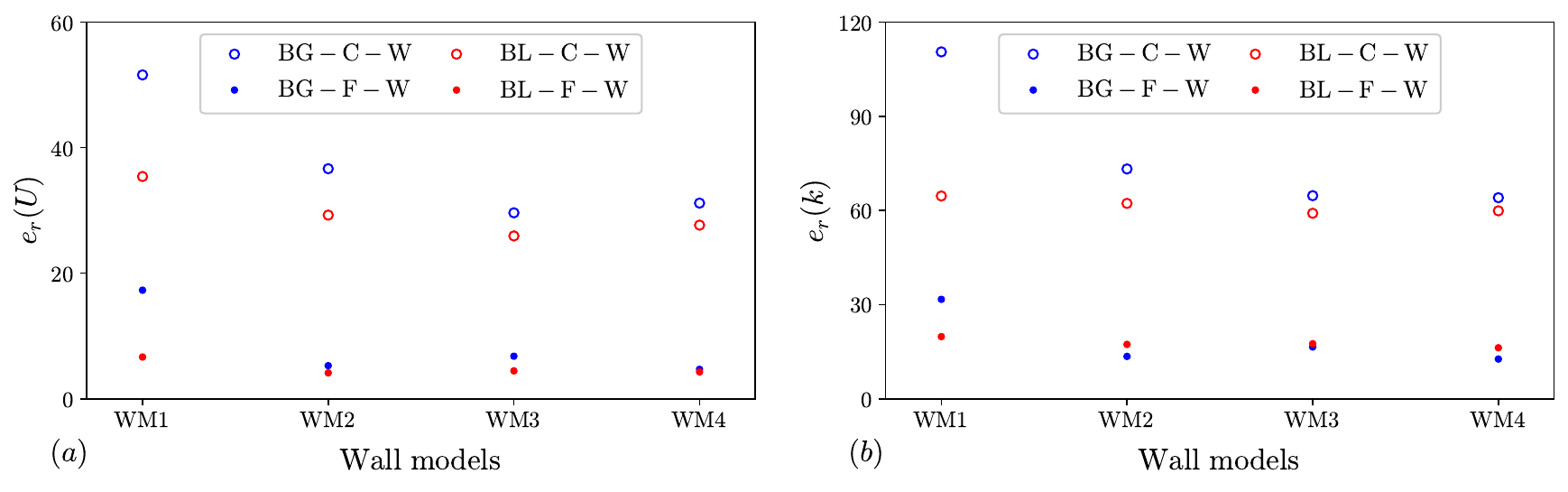}
    \caption{The percentage error for the mean velocity and TKE from WMLES of periodic hill flow at $Re_H = 10595$.
    The cases are BG-C/F-W and BL-C/F-W, which are summarized in Tables \ref{tab:overview_numerics} and \ref{tab:mesh_info}.
    The different wall models are summarized in Table \ref{tab:overview_wall_models}.}
    \label{fig:U_TKE_err_perHill_WALE}
\end{figure}

Similar to the channel flow cases, we summarize the errors for the mean velocity and TKE using the formulas in Eq.~\eqref{eq:errors_U_TKE}, including additional averaging along the streamwise direction.
Firstly, for the mean velocity, we see that the BL-C/F-W cases consistently perform better than the BG-C/F-W cases.
We note that this is the opposite conclusion of the channel flow cases.
Still, when non-equilibrium wall models are used, the difference between the BG-C/F-W and BL/C/F-W cases is less pronounced, especially on the fine mesh.
For the TKE, we first observe that the errors are much larger than for the mean velocity, as was also observed for the channel flow cases.
Further, we see that the BL-C/F-W performs better than BG-C/F-W when using the EWM (WM1); however, when non-equilibrium wall models are used, the BG-C/F-W and BL-C/F-W cases perform quite similar.

\begin{figure}[ht]
    \centering
    \includegraphics[width=0.9\textwidth]{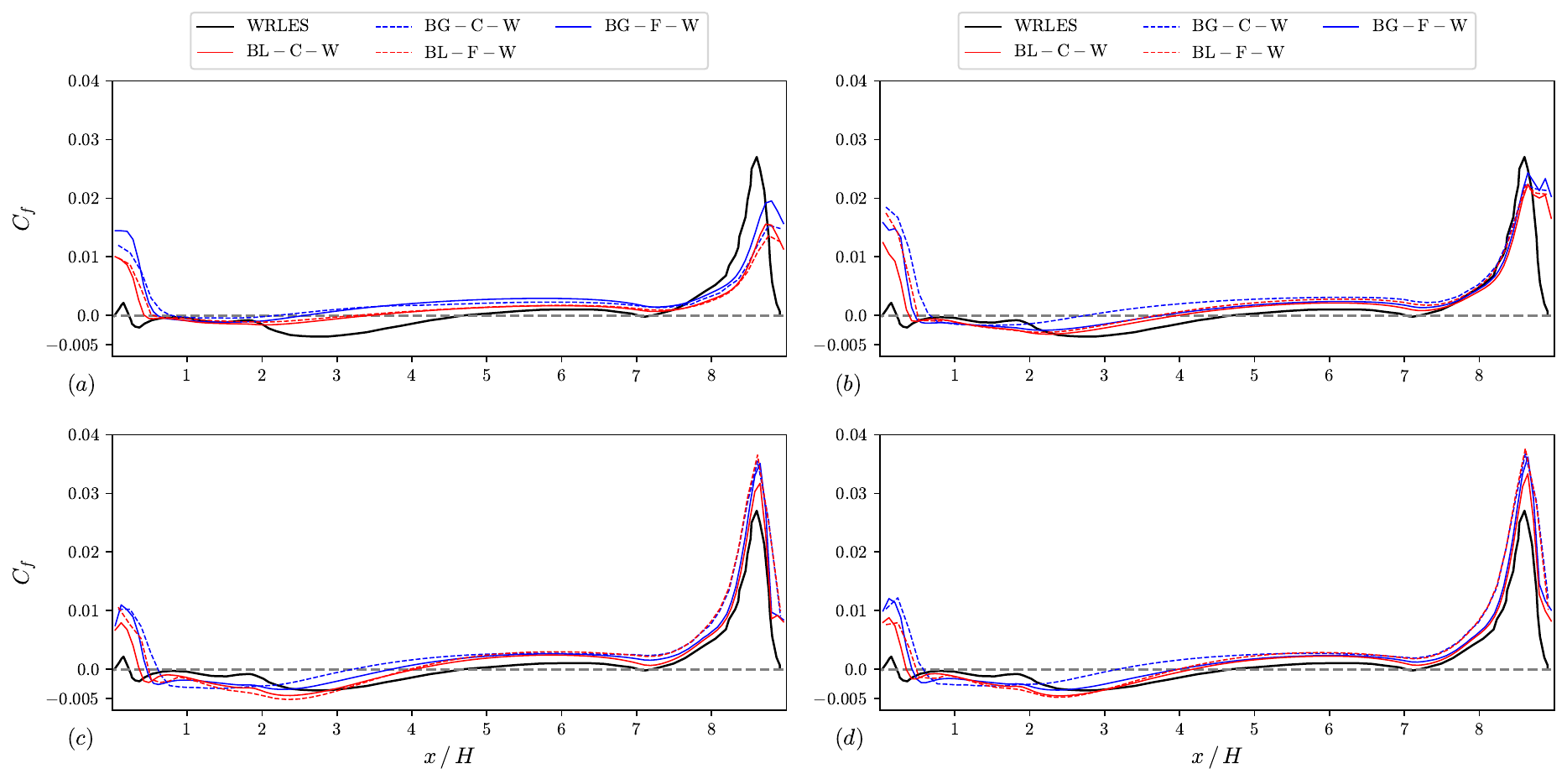}
    \caption{The skin friction coefficient $C_f$ from WMLES of turbulent flow over periodic hills.
    The cases are BG-C/F-W and BL-C/F-W, which are summarized in Tables \ref{tab:overview_numerics} and \ref{tab:mesh_info}.
    The wall models are: $(a)$ WM1, $(b)$ WM2, $(c)$ WM3, and $(d)$ WM4, see Table \ref{tab:overview_wall_models}.}
    \label{fig:Cf_perHill_WALE}
\end{figure}

We now consider the wall shear results presented in the form of the skin friction coefficient $C_f$, see Eq. \eqref{eq:skin_friction_coeff}.
The results are shown in Figure \ref{fig:Cf_perHill_WALE}.
From Figure \ref{fig:Cf_perHill_WALE} (a), we see that the EWM (WM1) barely captures any separation at all and that the $C_f$ value is significantly under-predicted at the front side of the hill.
There is also a streamwise shift in the position of the maximum $C_f$ value compared to the WRLES reference solution.
These observations are consistent with the behavior seen in the \textit{a-priori} analysis; see Figure \ref{fig:WM_tauw_results} (a).
Similarly, for WM2, WM3, and WM4, we also see a reasonable agreement between the \textit{a-priori} results in Figure \ref{fig:WM_tauw_results} and the \textit{a-posteriori} results in Figure \ref{fig:Cf_perHill_WALE}, however, the agreement is less pronounced than for the EWM (WM1).
Generally, we observe that the non-equilibrium wall models do a better job of capturing separation than the EWM (WM1).
Regarding the results on the coarse (C) and fine (F) meshes, we overall see more accurate predictions on the fine mesh.
Further, we note that all of the wall models struggle with predictions at the top of the hill where the convective term is most active and there is a strong imbalance between the convective and pressure gradient terms, see Figure \ref{fig:u_conv_vs_p_grad_near_wall}.
This suggests that a non-equilibrium wall model that includes both the convective and pressure gradient terms could improve the performance.

\begin{figure}[ht]
    \centering
    \includegraphics[width=0.5\textwidth]{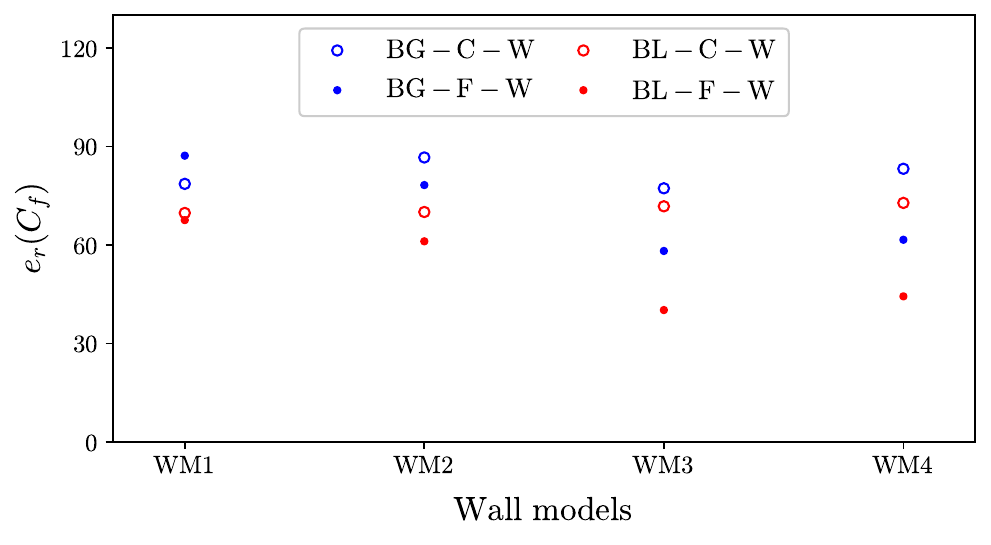}
    \caption{The percentage error for the mean friction coefficient from WMLES of periodic hill flow at $Re_H = 10595$.
    The cases are BG-C/F-W and BL-C/F-W, which are summarized in Tables \ref{tab:overview_numerics} and \ref{tab:mesh_info}.
    The different wall models are summarized in Table \ref{tab:overview_wall_models}.}
    \label{fig:tauw_err_perHill_WALE}
\end{figure}

To further quantify the predictive performance, we summarize the errors in the wall shear stress predictions using a similar formula as for the mean velocity and TKE, see Eq.~\eqref{eq:errors_U_TKE}.
The resulting errors are shown in Figure \ref{fig:tauw_err_perHill_WALE}.
First, we observe that the errors are large compared to both the mean velocity and TKE, see Figure \ref{fig:U_TKE_err_perHill_WALE}.
Second, we see that the BL-C/F-W cases outperform the BG-C/F-W cases for all four wall models.
Regarding the best performance, this is achieved with wall models WM3 and WM4.
This highlights that inclusion of the pressure gradient is beneficial in predicting the wall shear stress, at least for the periodic hill cases.
Still, given that even the smallest errors are quite large, there is clearly room for improvement.

\begin{figure}[ht]
    \centering
    \includegraphics[width=0.95\textwidth]{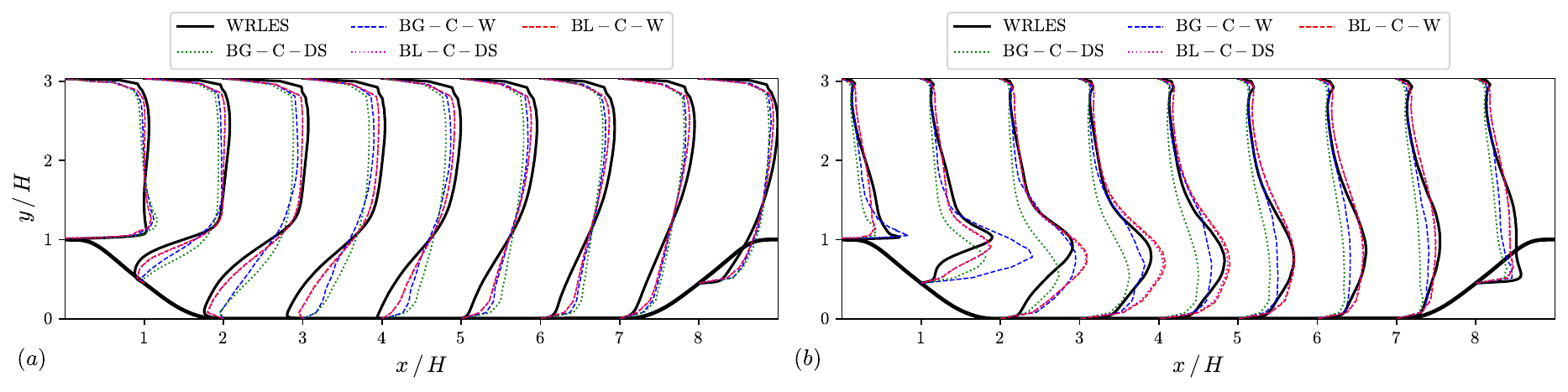}
    \caption{Results from WMLES of periodic hill flow at $Re_H = 10595$:
    (a) mean velocity and (b) mean TKE.
    The cases are BG-C-W/DS and BL-C-W/DS, see Tables \ref{tab:overview_numerics} and \ref{tab:mesh_info}, and the EWM (WM1) is used for wall modeling, see Table \ref{tab:overview_wall_models}.}
    \label{fig:U_perHill_SGSTests}
\end{figure}

Finally, to end this section, we consider the effect of using different SGS models.
For reasons of presentation, we only consider the coarse mesh cases, i.e., BG-C-DS and BL-C-DS, see Tables \ref{tab:overview_numerics} and \ref{tab:mesh_info}, using the EWM (WM1), and focus on the mean velocity and TKE.
As seen from Figure \ref{fig:U_perHill_SGSTests} (a), the mean velocity trends are the same as seen for the channel flow, see Figure \ref{fig:U_chan_SGSTests}.
Specifically, changes from using the two different SGS models are observed for the BG cases but not for the BL cases.
This is because the numerical dissipation introduced by the LUST scheme tends to overwhelm the contribution from the SGS model.
For the BG cases, using the DS model instead of the WALE model results in slightly worse mean velocity predictions over the majority of the domain.
On the other hand, for the TKE shown in Figure \ref{fig:U_perHill_SGSTests} (b), we see that the DS model gives improvements around $x / h \approx 1$ compared to the WALE model; however, in the remaining parts of the domain, the DS model gives worse predictions than the WALE model.
Taken together, we observe that the WALE model gives better predictions than the DS model for this case. 

\section{Conclusions}
\label{sec:conclusions}

In this work, we have carried out an OpenFOAM-based WMLES campaign covering turbulent flows in channels and over periodic hills.
The campaign aimed to determine the performance of the four different wall models summarized in Table \ref{tab:overview_wall_models} and to investigate the sensitivity of the results to changes in numerics, mesh resolution, and SGS modeling.
An overview of the cases, their details, and nomenclature are given in Tables \ref{tab:overview_numerics}, \ref{tab:mesh_info}, and \ref{tab:overview_wall_models}.
All four wall models are ODE wall models, which are based on solving simplified versions of the TBLEs Eq. \eqref{eq:TBLE}.
For WM1 and WM2 no non-equilibrium terms are included, i.e., $F_i = 0$, while WM3 and WM4 includes the pressure gradient term, i.e., $F_i = - \partial \langle p \rangle$.
Further, WM1 and WM3 use the van Driest eddy-viscosity Eq. \eqref{eq:van_Driest_eddy_viscosity} and WM2 and WM4 use the Duprat eddy-viscosity Eq. \eqref{eq:Duprat_eddy_viscosity}.
We note that WM1 is an EWM and serves as a baseline for comparison.
In terms of numerics, we mainly investigate the effect of using different discretizations of the convective term.
Specifically, we use the Gauss linear (BG cases) and LUST (BL cases) schemes, see Section \ref{sec:sim_details}.
The two different SGS models used are the WALE and DS models, see Section \ref{sec:LES}.
The main conclusions of the campaign are given below.
This is followed by a short discussion of their implications for future improvements.

For the channel flow, good results are obtained for the BG-C/F-W cases using all four wall models.
We note that wall models W2 and W4 are slightly more accurate than W1 and W3 in terms of the mean velocity profile.
The differences for the TKE, however, are not significant.
This indicates that the Duprat model Eq. \eqref{eq:Duprat_eddy_viscosity} is preferable to the van Driest model Eq. \eqref{eq:van_Driest_eddy_viscosity} in equilibrium flows.
On the other hand, the mean velocity and TKE from the BL-C/F-W cases show much larger errors than the BG-C/F-W cases.
To explain the over-prediction of the TKE in the first two off-wall cell centers, the production and SGS dissipation of TKE were examined.
We found that the BL cases have a lot less SGS dissipation and a slightly amplified production of TKE when compared with the BG cases, explaining the observed over-prediction.
For SGS modeling, comparing the WALE and DS models, we saw almost no changes in the mean velocity for both the BG and BL cases.
The TKE, on the other hand, shows clear variation between WALE and DS models for the BG cases, primarily at the second off-wall grid point.
The variation for the BL cases is very small, however, due to the additional dissipation introduced by the LUST scheme drowning out the SGS contribution.
Regarding the robustness of the results, the additional test cases in \ref{app:Add_num_tests} covering time step size, temporal discretization, and iterative solvers all showed only slight changes in both the mean velocity and TKE.
Based on these results, the following best practices for channel flow are proposed: Use the BG numerics, the WALE SGS model, and wall models WM2 or WM4.

The periodic hill case is less clear in terms of the best performance; however, some general trends do emerge.
One clear trend is that the non-equilibrium wall models WM2, WM3, and WM4 perform better in terms of mean velocity, TKE, and wall shear stress than the EWM (WM1) for both the BG-C/F-W and BL-C/F-W cases.
Another general trend is that the BL-C/F-W cases have better mean velocity results (especially in the separated region) than the BG-C/F-W cases.
This is especially pronounced on the coarse (C) mesh and when using the EWM (WM1).
The trend is similar for TKE results; however, for the non-equilibrium wall models on the fine (F) mesh, the BG cases show a small improvement over the BL cases.
In regards to the wall shear stress, the BL-C/F-W cases generally outperform the BG-C/F-W cases.
We also observe that the best performance is seen for wall models WM3 and WM4, which both include the pressure gradient.
For SGS modeling, comparing the WALE and DS models, we see some variation in both the mean velocity and TKE for the BG cases, with the WALE model giving better overall results than the DS model.
Similar to the channel flow, we see no discernible variation for the BL cases.
Further, as for the channel flow cases, we also considered additional test cases presented in \ref{app:Add_num_tests} covering time step size, temporal discretization, and iterative solvers.
These changes were found to have only a weak effect on the results.
Based on these results, the following best practices for periodic hill flow are proposed: Use the BL numerics, the WALE SGS model, and wall models WM3 or WM4.

As to future improvements, we focus the discussion on the two factors which was found to have the largest impact on the results.
These are the discretization of the convective term and wall modeling.
For the former, we observe that using the centered Gauss linear scheme gives better results in turbulent channel flow than using the upwind-biased LUST scheme.
For the periodic hill, the opposite was observed, with the LUST scheme having better overall performance than the Gauss linear scheme.
While this shows some benefit to introducing numerical dissipation in separated flows, it is clearly not universal across different flows, and thus, optimal use of this additional dissipation would need flow-specific tuning.
Additionally, the improvements observed for the LUST scheme are seen to shrink when more accurate wall models are used.
This suggests that using the Gauss linear scheme (or similar low-dissipation schemes) combined with improved wall modeling and a more dissipative SGS could lead to the best overall future performance.
When working with unstructured meshes, however, stability becomes a primary concern.
Still, in this case, numerical dissipation can be added locally where it is needed, e.g., using the approach from \citep{khalighi2011unstructured}. 
Moving on to wall modeling, the \textit{a-priori} investigations from Section \ref{subsubsec:a_priori_perhill} clearly indicated that additional improvement could be obtained by using a non-equilibrium wall model that includes both the convective and pressure gradient terms in a consistent manner.
As the wall modeling library from \citep{mukha2019library} already supports ODE wall models, the ODE-based non-equilibrium wall model from \citep{kamogawa2023ordinary} would be a natural choice to implement next. 

\section*{CRediT authorship contribution statement}

\textbf{Christoffer Hansen}: Writing – original draft, Investigation, Validation, Formal analysis, Visualization, Software, Data curation 
\textbf{Xiang I. A. Yang}: Writing – review \& editing, Methodology, Conceptualization.
\textbf{Mahdi Abkar}: Writing – review \& editing, Methodology, Conceptualization, Supervision, Project administration, Funding acquisition.

\section*{Declaration of competing interest}

The authors declare that they have no known competing financial interests or personal relationships that could have appeared to influence the work reported in this paper.

\section*{Acknowledgment}

\noindent C.H. and M.A. acknowledge the financial support from the Independent Research Fund Denmark (DFF) under Grant No. 1051-00015B.
X.Y. acknowledges financial support from the US Office of Naval Research contract N00014-24-1-2170 and Air Force Office of Scientific Research contract FA9550-23-1-0272.
This work was also partially supported by the Danish e-Infrastructure Cooperation (DeiC) National HPC under grant number DeiC-AU-N2-2023005.

\appendix

\section{Additional numerical tests}
\label{app:Add_num_tests}

\begingroup
\setlength{\tabcolsep}{6pt} 
\renewcommand{\arraystretch}{1.2} 
\begin{table}[ht]
\centering
\begin{tabular}{|c|c|c|c|c|c|}
\hline
             & \textbf{temporal}  & \textbf{divergence (conv. term)} & \textbf{U eq.} & \textbf{p eq.} & \textbf{CFL} \\ \hline
\textbf{TSG} & backward           & Gauss linear & PBiCGStab    & PBiCGStab & 0.2 \\ \hline
\textbf{TSL} & backward           & LUST         & PBiCGStab    & PBiCGStab & 0.2 \\ \hline
\textbf{TDG} & Crank-Nicolson 0.9 & Gauss linear                     & PBiCGStab      & PBiCGStab      & 0.4          \\ \hline
\textbf{TDL} & Crank-Nicolson 0.9 & LUST         & PBiCGStab    & PBiCGStab & 0.4 \\ \hline
\textbf{ITG} & backward           & Gauss linear & smoothSolver & GAMG      & 0.4 \\ \hline
\textbf{ITL} & backward           & LUST         & smoothSolver & GAMG      & 0.4 \\ \hline
\end{tabular}
\caption{Overview of numerics for the additional test cases.
The TSG and TSL cases consider a smaller time step size, the TDG and TDL cases consider a different temporal discretization, and the ITG and ITL cases consider different iterative solvers.}
\label{tab:overview_additional_tests}
\end{table}
\endgroup

We present supplementary WMLES cases to investigate the sensitivity of the results in Section \ref{sec:results}.
Two cases, TSG and TSL, are included to investigate the sensitivity to the timestep size.
The numerics are the same as the BG and BL cases, see Table \ref{tab:overview_numerics}, but with a maximum CFL number of 0.2.
The next two cases, TDG and TDL, investigate temporal discretization.
The new discretization is a blend of the implicit Crank-Nicolson scheme and the implicit Euler scheme.
The blending is done using a weighting factor, often chosen as 90\% Crank-Nicolson and 10\% Euler.
In OpenFOAM terminology, this is specified as "Crank-Nicolson 0.9".
The last two cases we include, ITG and ITL, explore the use of different iterative solvers for solving the momentum and pressure equations.
In the new cases, the momentum equations are solved using smoothSolver, while the pressure equation is solved using the Geometric-Algebraic Multi-Grid (GAMG) method.
Both smoothSolver and GAMG utilize a smoother, which we have chosen as Gauss-Seidel.
For additional details on the iterative solvers, see, e.g., \citep{FVM_OpenFOAM_book_2015}.
An overview of the six additional cases is given in Table \ref{tab:overview_additional_tests}.
Further, the meshing and SGS modeling are the same as in Table \ref{tab:mesh_info}.
Finally, for all of these additional cases, only simulations using EWM (WM1), see Table \ref{tab:overview_wall_models}, were carried out.

\subsection{Additional results for channel flow}

\begin{figure}[ht]
    \centering
    \includegraphics[width=0.75\textwidth]{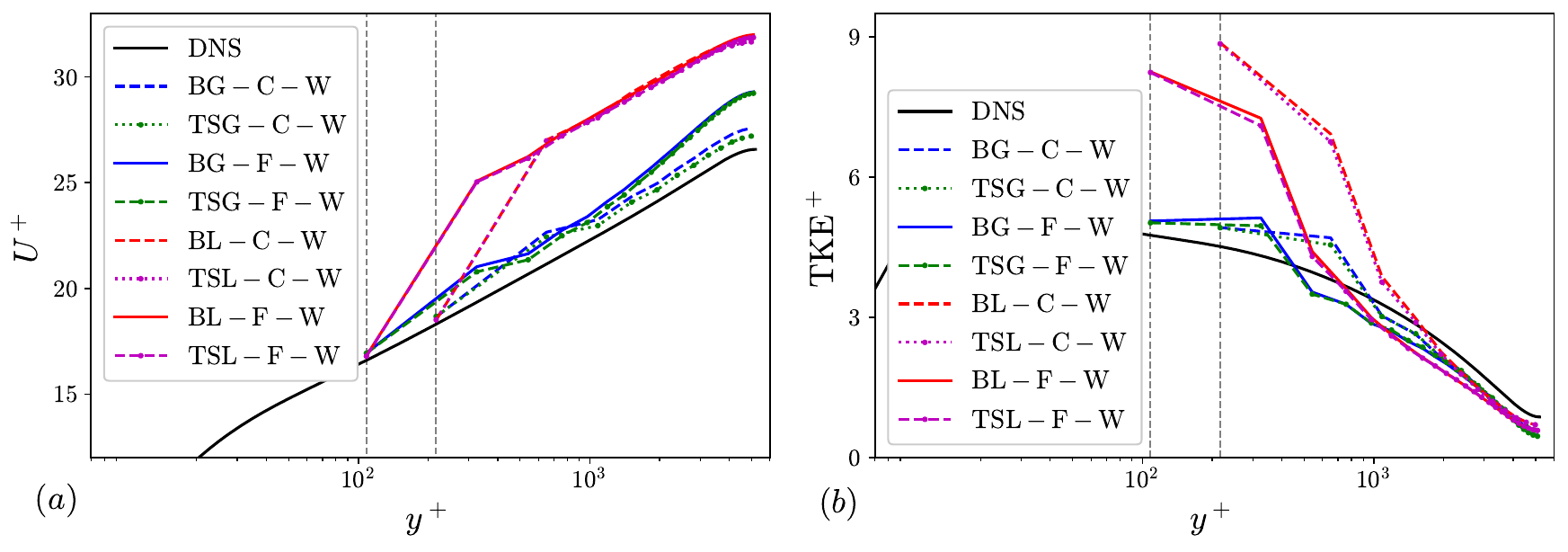}
    \caption{Time step tests from WMLES of channel flow at $Re_\tau = 5200$: (a) mean velocity and (b) TKE.
    The cases are BG-C/F-W and BL-C/F-W compared with TSG-C/F-W and TSL-C/F-W, see Tables \ref{tab:overview_numerics}, \ref{tab:mesh_info}, and \ref{tab:overview_additional_tests}.
    The wall model is the EWM (WM1), see Table \ref{tab:overview_wall_models}.}
    \label{fig:U_chan_timeStepTests}
\end{figure}

We consider the additional results for the turbulent channel flow.
We start by looking at the effect of decreasing the time step size by comparing the BG and BL cases with the TSG and TSL cases; see Tables \ref{tab:overview_numerics} and \ref{tab:overview_additional_tests}, respectively, for the details. 
From Figure \ref{fig:U_chan_timeStepTests}, we see that differences in both the mean velocity and TKE are very minor.
This shows that having a maximum CFL number of 0.4 is sufficient for WMLES of turbulent channel flow.

\begin{figure}[ht]
    \centering
    \includegraphics[width=0.75\textwidth]{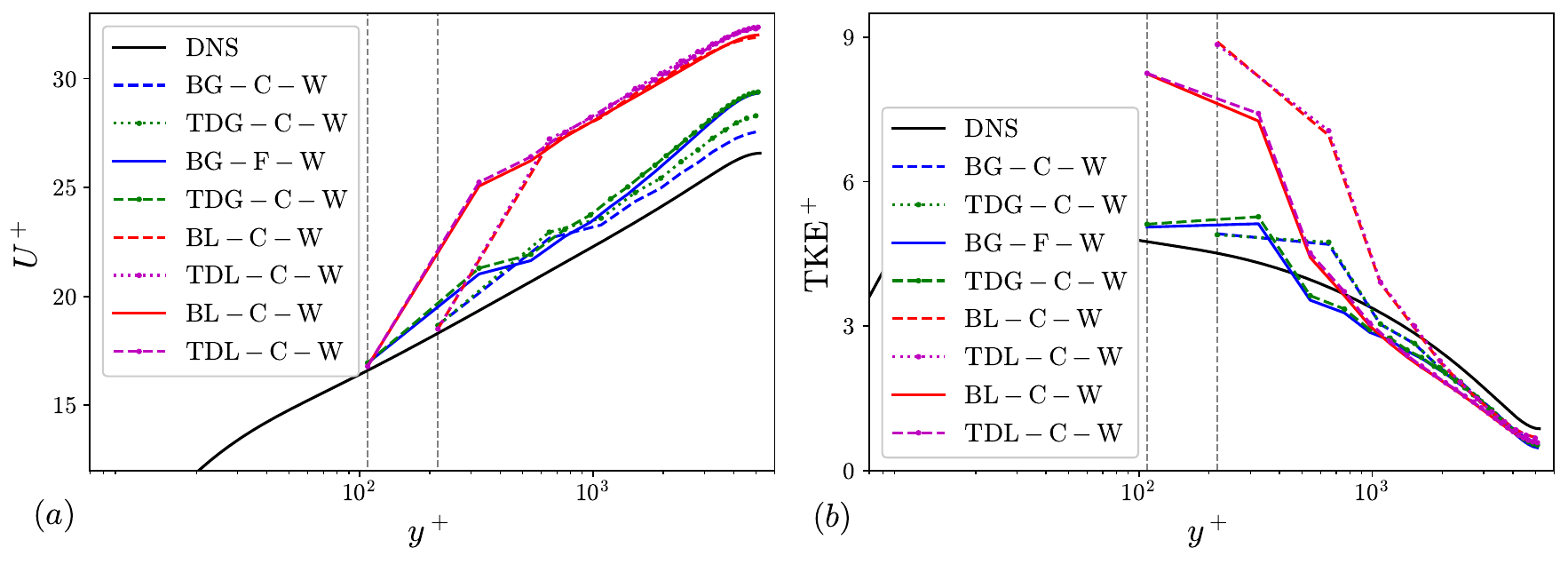}
    \caption{Time discretization tests from WMLES of channel flow at $Re_\tau = 5200$: (a) mean velocity and (b) TKE.
    The cases are BG-C/F-W and BL-C/F-W compared with TDG-C/F-W and TDL-C/F-W, see Tables \ref{tab:overview_numerics}, \ref{tab:mesh_info}, and \ref{tab:overview_additional_tests}.
    The wall model is the EWM (WM1), see Table \ref{tab:overview_wall_models}.}
    \label{fig:U_chan_timeDiscTests}
\end{figure}

Next, we consider the effect of using different time discretizations by comparing the BG and BL cases with the TDG and TDL cases, see Tables \ref{tab:overview_numerics} and \ref{tab:overview_additional_tests}.
Looking at the results in Figure \ref{fig:U_chan_timeDiscTests}, we see that there is little variation between the different cases.
The largest variation observed is for the mean velocity, see Figure \ref{fig:U_chan_timeDiscTests} (a), and is between the BG and TDG cases on the coarse mesh.
Based on these results, both time discretizations are reasonable choices for WMLES of turbulent channel flows.

\begin{figure}[ht]
    \centering
    \includegraphics[width=0.75\textwidth]{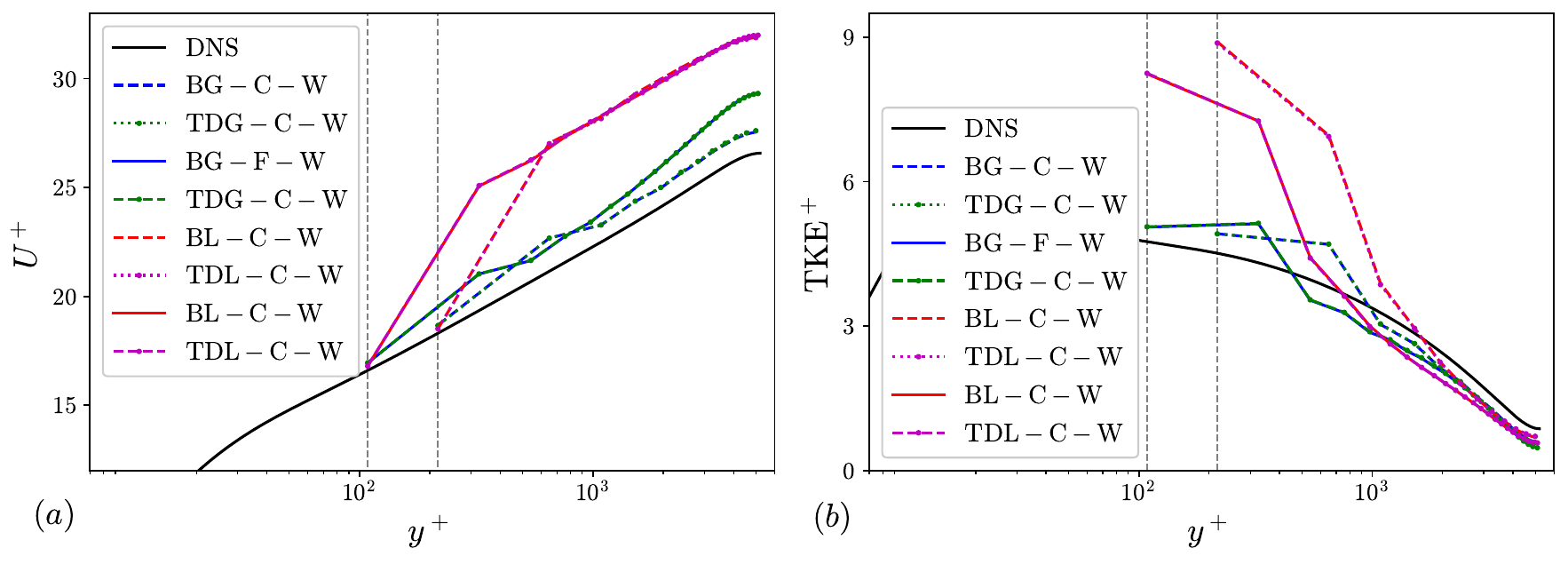}
    \caption{Iterative solver tests from WMLES of channel flow at $Re_\tau = 5200$: (a) mean velocity and (b) TKE.
    The cases are BG-C/F-W and BL-C/F-W compared with ITG-C/F-W and ITL-C/F-W, see Tables \ref{tab:overview_numerics}, \ref{tab:mesh_info}, and \ref{tab:overview_additional_tests}.
    The wall model is the EWM (WM1), see Table \ref{tab:overview_wall_models}.}
    \label{fig:U_chan_solverTests}
\end{figure}

Finally, we consider the effect of using different iterative solvers for solving the momentum and pressure equations. 
The comparison is between the BG and BL cases and the ITG and ITL cases, see Tables \ref{tab:overview_numerics} and \ref{tab:overview_additional_tests}.
From the results in Figure \ref{fig:U_chan_solverTests}, we see that there is no noticeable variation when using the different iterative solvers.
This shows that both choices for iterative solvers are appropriate for WMLES for turbulent channel flow.

\subsection{Additional results for periodic hill}

\begin{figure}[ht]
    \centering
    \includegraphics[width=0.98\textwidth]{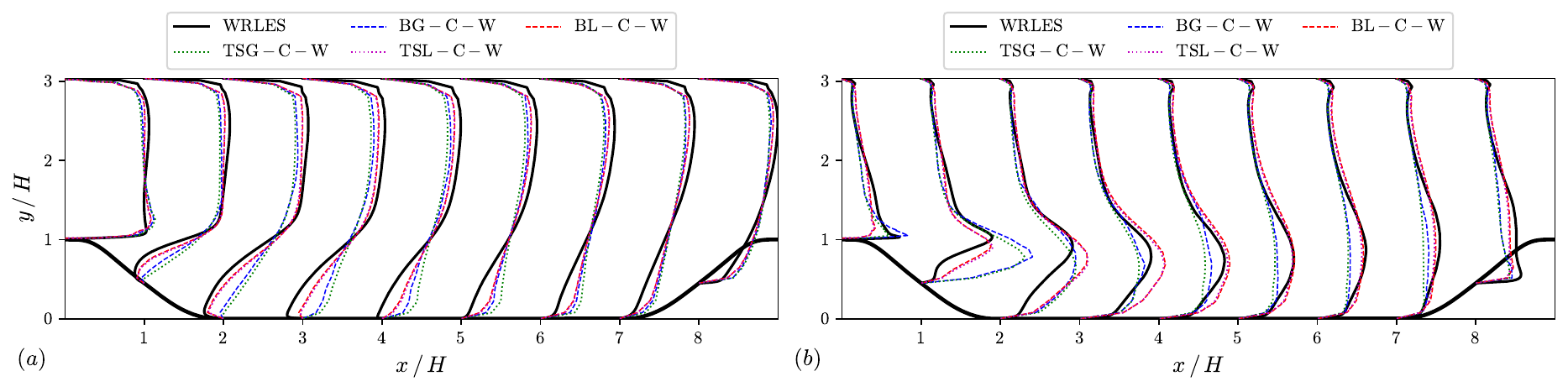}
    \caption{Time step tests from WMLES of periodic hill flow at $Re_H = 10595$: (a) mean velocity and (b) TKE.
    The cases are BG-C-W and BL-C-W compared with TSG-C-W and TSL-C-W, see Tables \ref{tab:overview_numerics}, \ref{tab:mesh_info}, and \ref{tab:overview_additional_tests}.
    The wall model is the EWM (WM1), see Table \ref{tab:overview_wall_models}.}
    \label{fig:U_perHill_timeStepTests}
\end{figure}

We now repeat the above sensitivity analysis for the periodic hill case.
The additional cases investigated are summarized in Table \ref{tab:overview_additional_tests}.
As for the channel flow, only the EWM (WM1) is considered here.
Further, we only consider results on the coarse mesh due to computational constraints.

First, we look at the effect of decreasing the time step size by comparing the BG and BL cases with the TSG and TSL cases.
The mean velocity and TKE results are shown in Figure \ref{fig:U_perHill_timeStepTests}.
We see some variation between the BG and TSG cases in both the mean velocity and TKE, with the BG case being slightly more accurate than the TSG case.
Still, the changes are quite small when compared to the difference seen between the BG and BL cases or between the different wall models in Section \ref{sec:results}.
For the BL and TSL cases, the changes are so small that they are barely visible.
We conclude that a maximum CFL number of 0.4 is reasonable for WMLES of turbulent flow over periodic hills.

\begin{figure}[ht]
    \centering
    \includegraphics[width=0.98\textwidth]{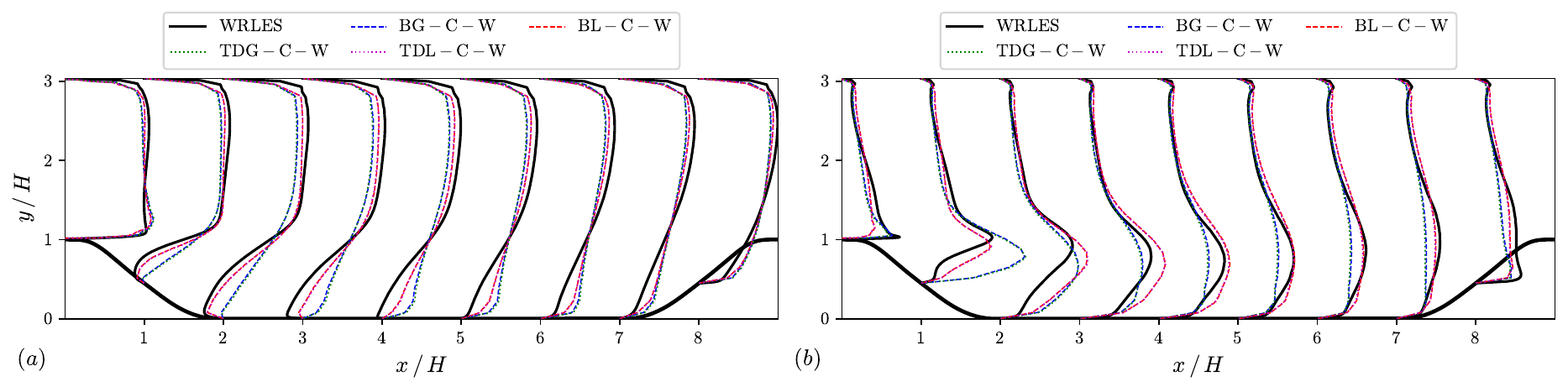}
    \caption{Time discretization tests from WMLES of periodic hill flow at $Re_H = 10595$: (a) mean velocity and (b) TKE.
    The cases are BG-C-W and BL-C-W compared with TDG-C-W and TDL-C-W, see Tables \ref{tab:overview_numerics}, \ref{tab:mesh_info}, and \ref{tab:overview_additional_tests}.
    The wall model is the EWM (WM1), see Table \ref{tab:overview_wall_models}.}
    \label{fig:U_perHill_timeDiscTests}
\end{figure}

Next, we consider the effect of using different time discretizations.
The BG and BL cases are compared with the TDG and TDL cases.
From Figure \ref{fig:U_perHill_timeDiscTests}, we see that no observable changes are present in the mean velocity or TKE from using different time discretizations.
This is contrary to the channel flow, where the time discretizations had a small but noticeable impact on the mean velocity between the BG and TDG cases, see Figure \ref{fig:U_chan_timeDiscTests} (a).

\begin{figure}[ht]
    \centering
    \includegraphics[width=0.98\textwidth]{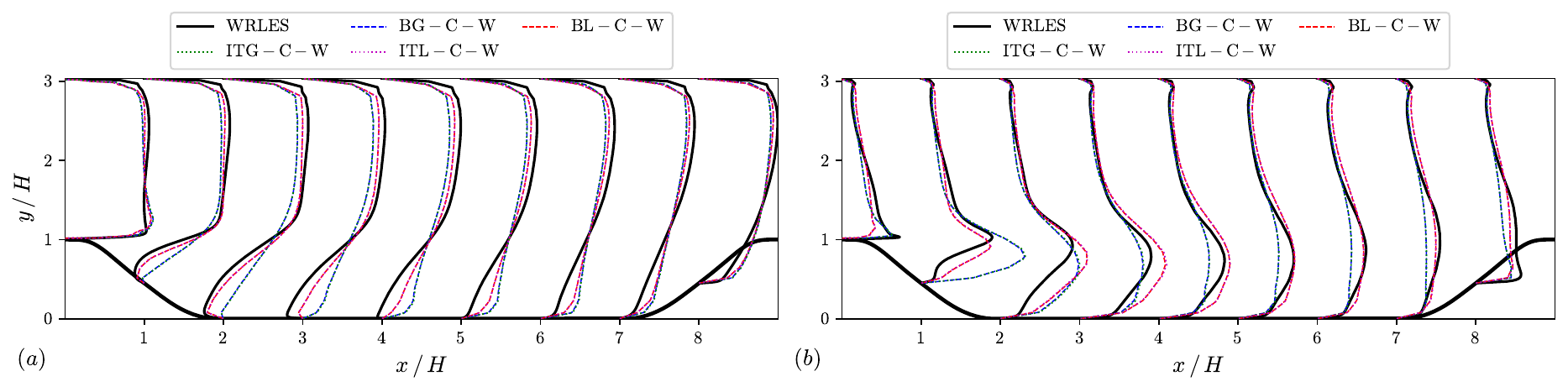}
    \caption{Iterative solver tests from WMLES of periodic hill flow at $Re_H = 10595$: (a) mean velocity and (b) TKE.
    The cases are BG-C-W and BL-C-W compared with ITG-C-W and ITL-C-W, see Tables \ref{tab:overview_numerics}, \ref{tab:mesh_info}, and \ref{tab:overview_additional_tests}.
    The wall model is the EWM (WM1), see Table \ref{tab:overview_wall_models}.}
    \label{fig:U_perHill_solverTests}
\end{figure}

Finally, we look at the effect of using different iterative solvers.
From Figure \ref{fig:U_perHill_solverTests}, it is clear that no discernible difference can be seen between the  mean velocity or TKE from using different iterative solvers.
This is consistent with the observations for the channel flow case, see Figure \ref{fig:U_chan_solverTests}.

\section{Wall-resolved results}
\label{app:WRLES_results}

We present the details and validation of the WRLES of the periodic hill case used as reference data for the \textit{a-priori} analysis in Section \ref{subsubsec:a_priori_perhill}.
The numerics are the same as the BG cases, see Table \ref{tab:overview_numerics}, and the SGS model model used in WALE.
The mesh is structured with hexahedral elements, similar to Figure \ref{fig:mesh_illustration} (b) but stretched towards the walls, and contains $N_x \times N_y \times N_z = 256 \times 128 \times 128$ elements along each coordinate direction.
This ensures elements that are almost isotropic in the streamwise-spanwise plane.
The mesh is stretched by a factor of 10 from the middle of the domain and toward the wall to ensure sufficient near-wall resolution.
This results in an inner scaled resolution which fulfills $\Delta x^+ , \Delta z^+ < 20$ and $\Delta y_1^+ < 2$ (except around $x/H \approx 8.6$ where the wall shear stress spikes and $\Delta y_1^+ \approx 2.5$ is reached), with $\Delta y_1$ the wall-normal distance to the first off-wall cell centers.
Thus, the current resolution aligns with common recommendations for WRLES; see, e.g., \citep{piomelli1996large}.

\begin{figure}[ht]
    \centering
    \includegraphics[width=0.98\textwidth]{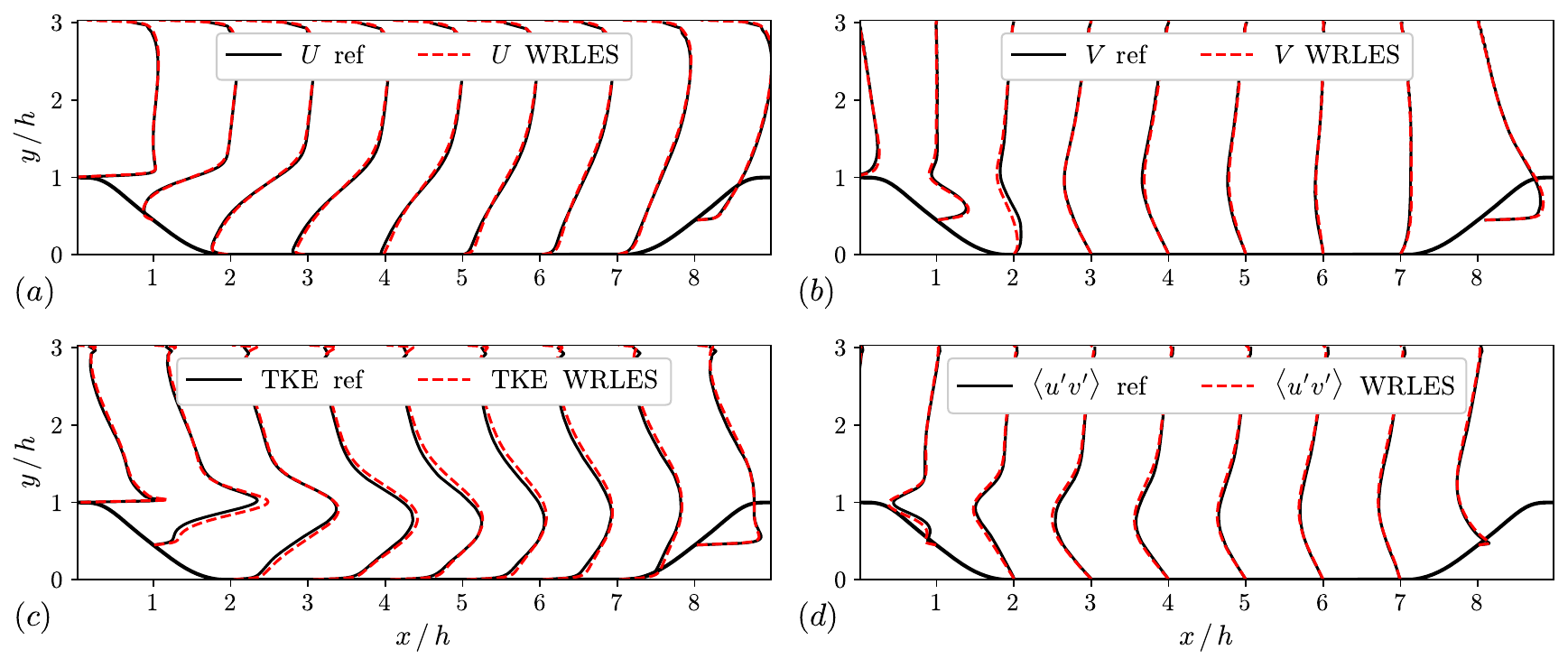}
    \caption{Comparison of WRLES results for the periodic hill case using OpenFOAM with reference WRLES data from \citep{frohlich2005highly}.}
    \label{fig:WRLES_vs_ref}
\end{figure}

To assess the accuracy of the WRLES results, we present a comparison with WRLES reference data from \citep{frohlich2005highly}.
Comparisons of the streamwise and wall-normal velocity components, the TKE, and the Reynolds shear stress are shown in Figure \ref{fig:WRLES_vs_ref}.
We see that the agreement between the two WRLES is generally excellent.
The most noticeable deviations are the underprediction of the wall-normal velocity at $x / H = 2$ and the TKE overshoot at $x / H = 1$. 
We note that the small discrepancies that are present between the current and reference simulations are of a similar magnitude as those observed between the two different solvers used in \citep{frohlich2005highly}.

\begin{figure}[ht]
    \centering
    \includegraphics[width=0.95\textwidth]{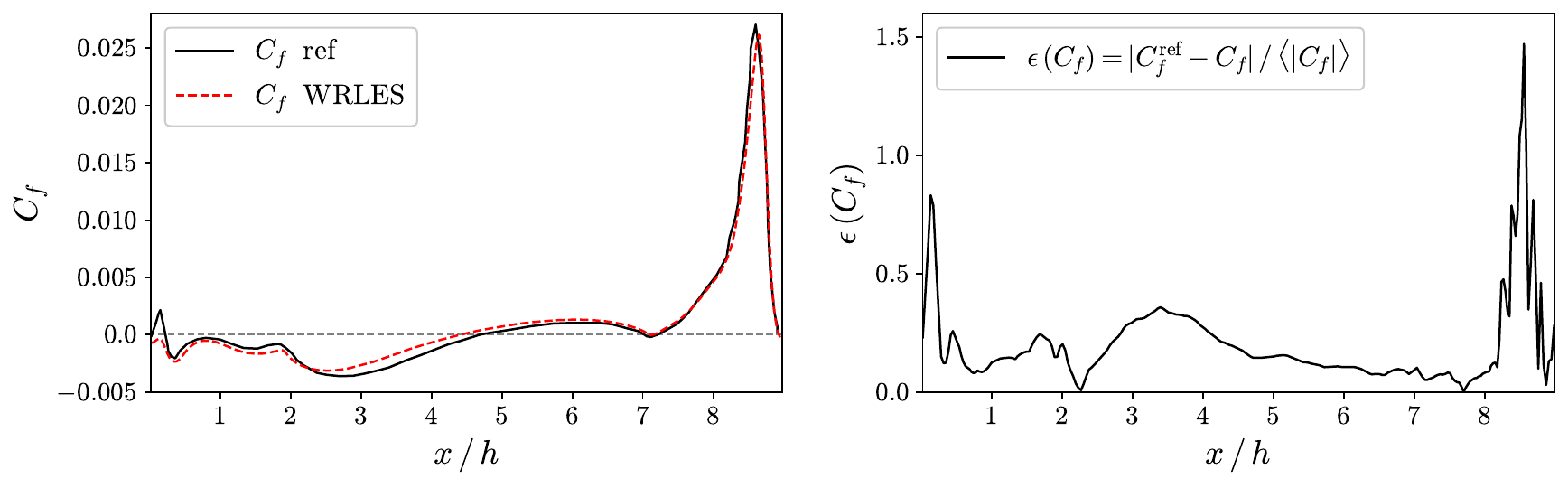}
    \caption{Comparison of $C_f$ from WRLES of the periodic hill case using OpenFOAM with reference WRLES data from \citep{frohlich2005highly}.}
    \label{fig:WRLES_tauw}
\end{figure}

Next, we consider the skin friction coefficient $C_f$, which is defined in Eq. \eqref{eq:skin_friction_coeff}.
Looking at Figure \ref{fig:WRLES_tauw} (a), we see that there is good agreement between the current and reference WRLES solutions.
To quantify the error further, we introduce the following percent-based relative error
\begin{align}
    \epsilon(C_f) = 100 \times \frac{|C_f^\text{ref} - C_f|}{\langle |C_f^\text{ref}| \rangle} .
\end{align}
Note that we normalize by the mean absolute value $\langle |C_f^\text{ref}| \rangle$, as normalization by $|C_f^\text{ref}|$ would lead to singularities at separation and reattachment points unless $|C_f^\text{ref} - C_f| = 0$ is fulfilled exactly.
From Figure \ref{fig:WRLES_tauw} (b), we see that the error remains comfortably within 2\% of the mean absolute value $\langle |C_f^\text{ref}| \rangle$.
This further highlights the accuracy of the current WRLES and its suitability for use as reference data in the \textit{a-priori} analysis in Section \ref{subsubsec:a_priori_perhill}.

\section*{Data availability}

Data will be made available on request.

\bibliographystyle{elsarticle-num-names} 
\bibliography{cas-refs}





\end{document}